\PassOptionsToPackage{pdfpagelabels=false}{hyperref}
\documentclass[useAMS,usenatbib]{mnras}
\pdfoutput=1
\usepackage[pdftex]{graphicx}
\usepackage{amsmath}
\usepackage{natbib}
\usepackage{geometry}

\usepackage{txfonts}
\usepackage{times}

\newcommand{\msun}{M$_{\odot}\,$}

\title[IllustrisTNG: Galaxy Colors]{\vspace{-1cm} First results from the IllustrisTNG simulations: \\ the galaxy color bimodality}
\author[D. Nelson et al.]{Dylan Nelson$^{1}$\thanks{E-mail: dnelson@mpa-garching.mpg.de},
Annalisa Pillepich$^{2}$, 
Volker Springel$^{3,4}$,
Rainer Weinberger$^{3}$, \newauthor
Lars Hernquist$^{5}$, 
R{\"u}diger Pakmor$^{3}$, 
Shy Genel$^{6}$,
Paul Torrey$^{7}$, \newauthor
Mark Vogelsberger$^{7}$, 
Guinevere Kauffmann$^{1}$,
Federico Marinacci$^{7}$,
Jill Naiman$^{5}$ \\\\
$^{1}$Max-Planck-Institut f\"{u}r Astrophysik, Karl-Schwarzschild-Str. 1, 85741 Garching, Germany\\
$^{2}$Max-Planck-Institut f\"{u}r Astronomie, K\"{o}nigstuhl 17, 69117 Heidelberg, Germany\\
$^{3}$Heidelberg Institute for Theoretical Studies, Schloss-Wolfsbrunnenweg 35, 69118 Heidelberg, Germany\\
$^{4}$Zentrum f\"{u}r Astronomie der Universit\"{a}t Heidelberg, ARI, M\"{o}nchhofstr. 12-14, 69120 Heidelberg, Germany\\
$^{5}$Harvard-Smithsonian Center for Astrophysics, 60 Garden Street, Cambridge, MA, 02138, USA\\
$^{6}$Center for Computational Astrophysics, Flatiron Institute, 162 Fifth Avenue, New York, NY 10010, USA\\
$^{7}$Kavli Institute for Astrophysics and Space Research, Department of Physics, MIT, Cambridge, MA, 02139, USA\\
}

\begin{document}

\maketitle

\begin{abstract}
We introduce the first two simulations of the IllustrisTNG project, a next generation of cosmological magnetohydrodynamical  
simulations, focusing on the optical colors of galaxies. We explore \textsc{TNG100}, a rerun of the original Illustris 
box, and \textsc{TNG300}, which includes 2$\times$2500$^3$ resolution elements in a volume twenty times larger. 
Here we present first results on the galaxy color bimodality at low redshift. Accounting for the attenuation of stellar light 
by dust, we compare the simulated (g-r) colors of $10^9$\,$<$\,M$_\star$/\msun$<$\,$10^{12.5}$ galaxies to the observed 
distribution from the Sloan Digital Sky Survey (SDSS). We find a striking improvement with respect to the original Illustris 
simulation, as well as excellent quantitative agreement with the observations, with a sharp transition in median 
color from blue to red at a characteristic $M_\star \sim 10^{10.5}$\,\msun\!. Investigating the build-up of the color-mass 
plane and the formation of the red sequence, we demonstrate that the primary driver of galaxy color transition 
is supermassive blackhole feedback in its low-accretion state. Across the entire population the median color transition 
timescale $\Delta t_{\rm green}$ is $\sim$\,1.6\,Gyr, a value which drops for increasingly massive galaxies. We find 
signatures of the physical process of quenching: at \textit{fixed} stellar mass, redder galaxies have lower SFRs, gas fractions, 
and gas metallicities; their stellar populations are also older and their large-scale interstellar magnetic fields weaker 
than in bluer galaxies. Finally, we measure the amount of stellar mass growth on the red sequence. Galaxies with \mbox{$M_\star$\,$>$\,$10^{11}$\,\msun\!} 
which redden at $z<1$ accumulate on average $\sim$\,25\% of their final $z=0$ mass post-reddening; at the same time, $\sim$\,18\% of 
such massive galaxies acquire half or more of their final stellar mass while on the red sequence. 
\end{abstract}



\section{Introduction}

Numerical simulations of cosmological galaxy formation have entered a new era. From the pioneering beginnings of dark-matter only 
simulations \citep{frenk83,davis85,navarro97}, through milestone projects of the past decade \citep{evrard02,spr05c,klypin11}, 
it has now become possible to also model the hydrodynamics of cosmic gas together with the principal baryonic physics driving 
the formation and evolution of galaxies. Increasingly sophisticated physical models are used to realize simulations of ever 
increasing scope. Larger volumes enable statistically robust comparisons of entire galaxy populations, while improved numerical 
methods combined with higher resolution realizations provide increasingly realistic results.
The emergence of cosmological hydrodynamical simulations as powerfully predictive theoretical models is embodied in the 
recent Illustris \citep{vog14b,vog14a,genel14,sijacki15} and EAGLE \citep{schaye15,crain15} simulations. In concert with other 
large-volume efforts \citep{dubois16,dolag16,dave16} these projects have convincingly demonstrated that hydrodynamical 
simulations of structure formation at kilo-parsec spatial resolution can reasonably reproduce the fundamental properties and 
scaling relations of observed galaxies. This zeroth order agreement has enabled many fruitful investigations; it has, at the 
same time, revealed many informative shortcomings.

Here we present IllustrisTNG, a simulation campaign which (i) retains the fundamental approach and physical model flavor of 
Illustris, (ii) alleviates many Illustris model deficiencies with respect to benchmark observations, and (iii) 
significantly expands the scope with simulations of larger volumes, at higher resolution, and with new physics. In doing so 
we continue to heavily rely on sub-resolution models for physics below scales of a few hundred parsecs: by definition, this 
includes the process of star formation, the details of supernovae feedback, and the growth and evolution of supermassive 
blackholes.

In order to improve upon our existing physical model, one of our first tasks is therefore to address regimes in which it 
fails to provide a reasonable match to observational data. To this end we identified a set of clear model deficiencies in the 
original Illustris simulation and specifically addressed them in the new TNG model \citep{pillepich17a,weinberger17}. 
The principal model changes are revised treatments of both stellar feedback driven galactic winds and supermassive blackhole feedback.
In this new context, we now investigate a key issue which existed in the previous simulations: the lack of a clear color 
bimodality between two distinct red and blue galaxy populations at low redshift. 
In \cite{vog14b} we presented the distributions of intrinsic (u-i) and (g-r) colors of simulated Illustris galaxies. 
Broadly, we concluded that the red peak for quiescent galaxies was either entirely absent, or not sufficiently pronounced, 
depending on stellar mass, while the transition from predominantly blue to red systems occurred at too large $M_\star$, and  
evidence for a bimodal color distribution at any mass was weak. We principally attributed this failure to the inability of 
Illustris blackhole feedback to entirely quench residual star formation in massive halos.

There are surprisingly few other investigations of color across the full galaxy mass range in cosmological hydrodynamical 
simulations. Analyzing the EAGLE simulation, \cite{trayford15} is by far the most comprehensive, demonstrating favorable 
comparison to low-redshift color-magnitude and luminosity function GAMA results, although this involves a combination 
of the fiducial volume with smaller box, higher resolution, re-calibrated models; despite good qualitative agreement, 
the coherent level of quantitative agreement is by no means perfect. In \cite{trayford16} the authors explore the 
evolution of the color-mass space, making the connection to both environmental-driven and blackhole feedback-driven color 
transition. Previously, \cite{cen14} mainly studied the reddening of satellite galaxies due to hydrodynamic processes, 
without including blackholes or comparing to observed color distributions. \cite{gabor12} used a phenomenological halo 
quenching model to study the formation of the red sequence, likewise omitting blackholes and attributing both central 
and satellite quenching to environments dominated by virialized gas. An update in \cite{dave17} where the enforced 
halo quenching mass varies with redshift also produces red central galaxies at the expense of excessively red satellites.

The emergence of a color bimodality was quickly identified as a critical challenge for semi-analytical models of galaxy 
formation as well \citep{cole00}. Specifically, models which introduced a central heating process from a supermassive 
blackhole were then able to recover the red colors of the most massive galaxies \citep{kang05,menci05,croton06,cattaneo07}. 
The co-evolution of a central blackhole and its host galaxy is an important ingredient \citep{bower06,somerville08,benson12}, 
and more recent efforts continue to explore galaxy color as a fundamental observational constraint 
\citep[e.g.][]{henriques15,zu17}.

Observationally, the color of a galaxy is a fundamental distillation of the emergent spectral distribution of light emitted 
by its constituent stellar populations. Measurements of galaxy colors are one of the most direct probes of the physical 
characteristics of extragalactic stars, with a correspondingly rich observational history. Modern large-scale surveys -- 
namely, the Sloan Digital Sky Survey (SDSS) -- enabled the discovery \citep{strateva01} and early characterization 
\citep{kauffmann03,blanton03} of the bimodal distribution of galaxy colors at low redshift. Based on visual classification, 
\cite{strateva01} demonstrated the rough relationship between color and morphology: that members of the red population are 
generally more spheroidal, while members of the blue population are more disk-like. \cite{baldry04} modeled the color 
distribution of SDSS galaxies using a double-Gaussian formalism, measuring the separate red and blue luminosity and mass 
functions. Beyond the local universe, \cite{bell04} confirmed the dual population picture at $0.2\,<\,z\,<\,1.0$ using the 
COMBO-17 survey, constraining the evolution of the red population. Subsequently, \cite{faber07} used COMBO-17 and DEEP2 to 
contrast the $z\,=\,0$ and $z\,=\,1$ red and blue galaxy populations, constraining the evolution of galaxies between, and 
within, each. More recently, \cite{taylor15} presented a sophisticated modeling of the overlapping red and blue galaxy 
populations from $z\,<\,0.1$ results of the GAMA survey, quantifying their respective properties and highlighting the care 
needed in their separation.

Beyond the fundamental relation of galaxy color and stellar mass, observations have also explored correlations between, 
and implications arising from, other galactic properties.
Studying environment, \cite{kauffmann04} and \cite{baldry06} used SDSS to 
demonstrate that while the location of the two populations in the D4000/color-mass plane does not correlate with (projected 
neighbor density) environment, the relative fraction in each population does. Thereafter \cite{peng10} advocated for the 
separability of mass and environmental (i.e. satellite) driven color transformation to $z\,=\,1$, quantifying an increasing 
red fraction with overdensity at fixed stellar mass.
Considering structural properties, \cite{kauffmann03b} found that stellar surface mass density and stellar concentration 
vary strongly with mass, separating at a characteristic threshold of $M_\star$\,$\sim$\,3$\times$10$^{10}$\,\msun\! likely 
established by baryonic feedback processes.
Focusing on morphology, \cite{schawinski14} used Galaxy 
Zoo classifications and found two distinct transition pathways, with different physical timescales, correlated with early 
versus late-type morphology. Leveraging the CANDELS survey up to $z\,=\,3$, \cite{pandya16} measured the prevalence of the 
intermediary population, as well as the occupation timescales and physical properties of its members. Recently, \cite{powell17} 
measured the color-mass distributions at $z\,=\,1$ separated by morphological type and argued for two distinct timescales 
and the possible role of blackhole feedback induced transformation.

In this paper we have two goals: to present the new TNG simulations, and to quantitatively assess their agreement with the 
$z\,=\,0$ observed color distribution, as a benchmark for improvement over the original Illustris simulation. 
Given this newly accessible theoretical regime, we then proceed to study the characteristics and 
origin of the galaxy color bimodality. In Section~\ref{sec_sims} we describe the simulations and their methods, while Section~
\ref{sec_color_model} details our modeling of galaxy colors and the effects of dust. Section~\ref{sec_obs_comp} compares the 
results of TNG to the observational data. In Section~\ref{sec_theory_interp} we provide a theoretical picture of the 
color transition process and the origin and buildup of the galaxy color bimodality. We conclude with a discussion in Section~
\ref{sec_discussion} and summarize with Section~\ref{sec_summary}. Finally, Appendix~\ref{sec_appendix} evaluates numerical 
convergence and details of our color modeling.


\section{The TNG Simulations} \label{sec_sims}

The IllustrisTNG project\footnote{\url{http://www.tng-project.org}} is the successor of the Illustris simulation 
\citep{vog14b,vog14a,genel14,sijacki15}. It uses an updated `next generation' galaxy formation model which includes 
both new physics as well as refinements to the original Illustris model. The complete description of the TNG galaxy formation 
model, including details of the physical processes we include and explorations of the model and model variations on smaller 
test volumes are presented in the two TNG methods papers \citep{weinberger17,pillepich17a}. For clarity, we avoid an 
exhaustive repetition of the details and here enumerate only the principal features.

TNG is a series of large, cosmological, gravo-magneto-hydrodynamical simulations incorporating a comprehensive model for galaxy 
formation physics. We use the \textsc{Arepo} code \citep{spr10} to solve the coupled equations of self-gravity and ideal, 
continuum magneto-hydrodynamics \citep[MHD;][]{pakmor11,pakmor13}, the former computed with the Tree-PM approach and the latter 
employing a Godunov/finite-volume method with a spatial discretization based on an unstructured, moving, Voronoi tessellation 
of the domain. The scheme is quasi-Lagrangian, second order in both space and time, uses individual particle time-stepping, and 
has been designed to efficiently execute large, parallel astrophysical simulations on modern supercomputer architectures -- 
our largest run used 24000 compute cores.

On top of this numerical framework the TNG simulations implement models for the key physical processes relevant for galaxy 
formation and evolution. These are: (i) microphysical gas radiative mechanisms, including primordial and metal-line cooling 
and heating with an evolving ultraviolet background field, (ii) star formation in the dense interstellar medium, (iii) stellar 
population evolution and chemical enrichment following supernovae Ia, II, and AGB stars, individually tracking H, He, C, N, 
O, Ne, Mg, Si, and Fe, (iv) stellar feedback driven galactic-scale outflows, (v) the formation, merging, and accretion of 
supermassive blackholes, (vi) multi-mode blackhole feedback operating in a thermal `quasar' mode at high accretion states 
and a kinetic `wind' mode at low accretion states. 

We describe the complete model used to run the TNG cosmological simulations in \cite{pillepich17a} -- see Table 1 of that work 
for the salient differences with respect to the original Illustris model -- together with \cite{weinberger17}, focusing on the 
blackholes and the high-mass end. Every aspect of the physical model and detail of its numerical implementation, including all 
parameter values and the simulation code version, are documented therein and \textit{entirely unchanged} for the simulations 
we present here. 

{\renewcommand{\arraystretch}{1.2}
\begin{table}
  \caption{Most important details of the two primary simulations presented here: TNG100 and TNG300. These are: the simulated 
           volume and box side-length (both comoving), the number of initial gas cells, dark matter particles, and Monte 
           Carlo tracers. The mean baryon and dark matter particle mass resolutions, in solar masses. The minimum allowed 
           adaptive gravitational softening length for gas cells (comoving Plummer equivalent), and the redshift zero softening 
           of the dark matter and stellar components in physical kilo-parsecs. 
           In Tables \ref{simTableBig} and \ref{simTableDM} we provide additional details, including the resolution series 
           and dark-matter only analogs of each box.}
  \label{simTable}
  \begin{center}
    \begin{tabular}{lcll}
     \hline\hline
     
 Run Name & & TNG100 & TNG300 \\ \hline
 Volume & [\,Mpc$^3$\,] & $110.7^3$ & $302.6^3$ \\
 $L_{\rm box}$ & [\,Mpc/$h$\,] & 75 & 205 \\
 $N_{\rm GAS}$ & - & $1820^3$ & $2500^3$ \\
 $N_{\rm DM}$ & - & $1820^3$ & $2500^3$ \\
 $N_{\rm TRACER}$ & - & $2 \times 1820^3$ & $1 \times 2500^3$ \\
 $m_{\rm baryon}$ & [\,M$_\odot$\,] & $1.4 \times 10^6$ & $1.1 \times 10^7$ \\
 $m_{\rm DM}$ & [\,M$_\odot$\,] & $7.5 \times 10^6$ & $5.9 \times 10^7$ \\
 $\epsilon_{\rm gas,min}$ & [\,pc\,] & 185 & 370 \\
 $\epsilon_{\rm DM,stars}^{z=0}$ & [\,kpc\,] & 0.74 & 1.48 \\
 $\epsilon_{\rm DM,stars}$ & [\,ckpc/$h$\,] & 1.0 $\rightarrow$ 0.5 & 2.0 $\rightarrow$ 1.0 \\
 \hline
 
    \end{tabular}
  \end{center}
\end{table}}

The TNG project is made up of three simulation volumes: TNG50, TNG100, and TNG300. Here we present the two larger volumes: 
the first simulation, \textbf{TNG100}, includes 2$\times$1820$^3$ resolution elements in a $\sim$\,100 Mpc (comoving) box. The 
baryon mass resolution is $1.4 \times 10^6$\,\msun\!, the gravitational softening length of the dark matter and stars is 0.7 
kpc at $z$\,=\,0, and the gas component has an adaptive softening with a minimum of 185 comoving parsecs. The phases are the 
same as original Illustris, enabling object-by-object comparison. The second simulation, \textbf{TNG300}, includes 
2$\times$2500$^3$ resolution elements in a $\sim$\,300 Mpc box, a volume $\sim$\,20 times larger. Its baryon mass resolution, 
collisionless softening, and gas minimum softening are $1.1 \times 10^7$\,\msun\!, 1.5 kpc, and 370 parsecs, respectively. 
The key details of each volume are given in Table \ref{simTable}. 
We also run two lower resolution realizations for each volume and dark-matter only analogs of every configuration, a total of 
twelve simulations which are detailed, and studied, in Appendix \ref{sec_appendix}. The high-resolution \textbf{TNG50} run 
will be introduced in future work (\textcolor{blue}{Nelson et al. in prep}, \textcolor{blue}{Pillepich et al. in prep}). 
It includes 2$\times$2160$^3$ resolution elements in a $\sim$\,50 Mpc box, with a baryon mass resolution of 
$8.5 \times 10^4$\,\msun\!, $z=0$ collisionless softening of 280 parsecs, and a minimum gas softening of 74 parsecs.

This is one in a series of five papers that introduce IllustrisTNG, each of the other four with a distinct scientific focus. 
In \textcolor{blue}{Naiman et al. (2017)} we study neutron star-neutron star mergers and r-process chemical enrichment, 
and in \textcolor{blue}{Pillepich et al. (2017)} we decompose the different stellar components at group and cluster scales and 
investigate the universality of stellar mass profiles.
In \textcolor{blue}{Marinacci et al. (2017)} we measure radio synchrotron emission from massive galaxy clusters, while 
in \textcolor{blue}{Springel et al. (2017)} we examine the spatial clustering of matter and of galaxies.

For TNG we adopt a cosmology consistent with recent constraints \citep{planck2015_xiii}, namely 
$\Omega_{\Lambda,0}=0.6911$, $\Omega_{m,0}=0.3089$, $\Omega_{b,0}=0.0486$, $\sigma_8=0.8159$, $n_s=0.9667$ and $h=0.6774$. 
Herein we identify gravitationally bound substructures using the \textsc{Subfind} algorithm \citep{spr01} 
and link them through time with the \textsc{SubLink} merger tree algorithm \citep{rodriguezgomez15}.


\section{Modeling Galaxy Colors} \label{sec_color_model}

We consider three models for computing the colors of galaxies: (A), (B), and (C). Each builds 
off the previous with increasing complexity. Model (A) sums up the intrinsic emission of each 
star particle in a galaxy, with no attenuation. Model (B) adds `unresolved' dust attenuation due
to dense gas birth clouds surrounding young stellar populations. Model (C) adds `resolved' dust 
attenuation due to the simulated distribution of neutral gas and metals. Therefore, the first 
two models are invariant to orientation, while colors from the third depend on viewing angle.
In this paper \textbf{all results use model (C) exclusively}, while the three models are compared 
in the appendix.

\subsection{Model (A) - no dust}

Each star particle in the simulation is modeled as a single-burst simple stellar population (SSP), whose 
birth time, metallicity, and mass are recorded. We use the FSPS stellar population synthesis code 
\citep{conroy09,conroy10,dfm14} with the Padova isochrones, MILES stellar library, and assuming a 
Chabrier IMF. The isochrones contain 22 metallicity steps from $-3.7$ to $-1.5$ (log total metal mass ratio), 
and 94 age steps from $-3.5$ to $1.15$ (log Gyr). For each point on this 2D grid, we convolve the resulting 
population spectrum with each of the observed pass-bands (SDSS u,g,r,i,z airmass 1.3), saving the total 
magnitude in each. Default models for dust emission and nebular line and continuum emission are included.

For each star particle we then use bicubic interpolation in this $(Z_i,t_{\rm ssp})$ plane to obtain $m_1$, 
the band magnitude of a 1\,\msun equivalent population. Multiplying by the initial mass $M_i$ gives 
the total band magnitude of the actual population $m = m_1 - 2.5 * \log_{10}(M_i / \rm{M}_\odot)$. For the total 
magnitude of each galaxy, we sum all the subhalo member star particles, with no additional restrictions. 
Galaxy colors are the difference between two of these total band magnitudes.

\subsection{Model (B) - unresolved dust}

Following \cite{cf00} we add a simple powerlaw extinction model for the attenuation by finite lifetime 
birth clouds surrounding young stellar populations as well as the ambient diffuse ISM. The intrinsic 
luminosity $L_{\rm i}(\lambda)$ of each stellar population is attenuated as 
$L_{\rm obs}(\lambda) / L_{\rm i}(\lambda) = e^{-\tau^{\rm B}_\lambda}$ where the wavelength dependent 
optical depth $\tau^{\rm B}_\lambda$ is then given as

\begin{equation}
\tau^{\rm B}_\lambda = 
  \left\{ \quad
    \begin{aligned}
      \tau_1(\lambda / \lambda_0)^{-\alpha_1} \quad,\, t_{\rm ssp} \le t_{\rm age} \quad \\
      \tau_2(\lambda / \lambda_0)^{-\alpha_2} \quad,\, t_{\rm ssp} > t_{\rm age} \quad
    \end{aligned}
  \right\}.
\end{equation}

\noindent All parameters are unchanged from their commonly adopted values. We take 
$\tau_1 = 1.0$, 
$\tau_2 = 0.3$ \citep[although see][]{wild11}, 
$\alpha_1 = \alpha_2 = 0.7$, 
$t_{\rm age} = 10$\,Myr, 
and $\lambda_0 = 550$\,nm. 
The resulting colors remain independent of viewing angle.

\subsection{Model (C) - resolved dust}

\begin{figure*}
\centerline{\includegraphics[angle=0,width=7.0in]{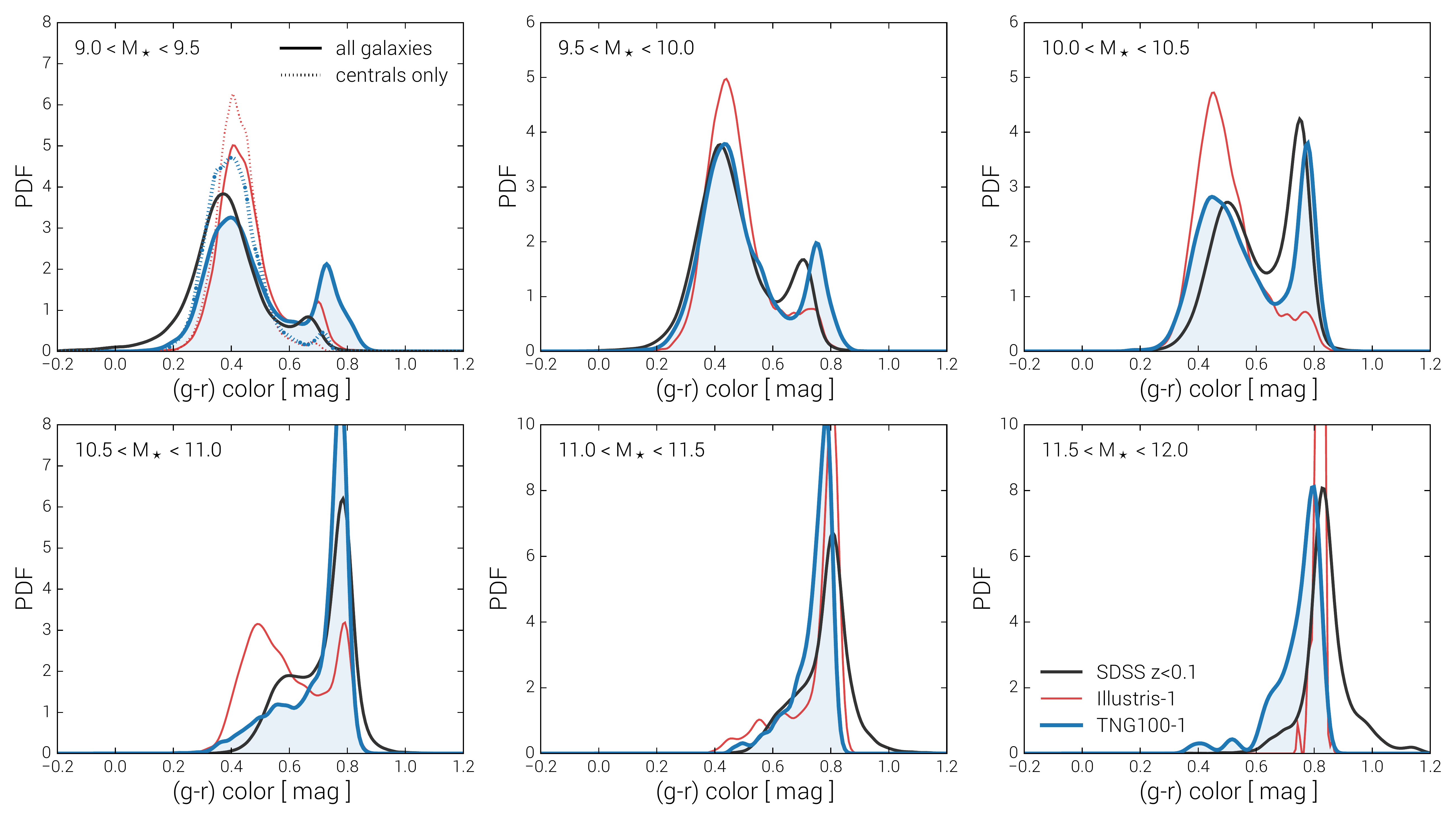}}
\caption{ The distribution of simulated versus observed (g-r) colors. In black the SDSS $z$\,$<$0.1 sample is shown, while red shows 
the result from the old Illustris simulation, and blue the result from our new TNG100 simulation of equivalent box size and 
resolution. In all cases, galaxies with stellar masses from $10^{9}$ to $10^{12}$\,\msun are included, divided into six mass bins as 
indicated in the six panels (log M$_\star$ values shown). One-dimensional kernel density weighted PDFs are shown (with a half-Scott 
bandwidth of $0.5N^{-1/5}$). In each case galaxies 
are shown regardless of if they are centrals or satellites. In addition, in the lowest mass panel we also decompose the simulated 
results and show centrals only (dotted lines), demonstrating that the somewhat excessive red peak at these masses arises from the 
red satellite population. Illustris and TNG100 galaxies are treated with the same dust model C (see Section \ref{sec_color_model}).
Evident in both SDSS and TNG100: the dominant blue population at low masses, the emergence of the red peak 
and the strong bimodality around $10^{10.5}$\,\msun\!, and the dropout of blue systems with the red population dominant at high masses.
 \label{fig_colors_1d_histos}} 
\end{figure*}

We follow the distribution of neutral gas in and around each simulated galaxy. Using its metal content 
we can then model the subsequent dust attenuation of the optical light 
emitted by stars. We reiterate that this obscuration is in addition to the previously described Model (B) component. 
Following \cite{guo09} and \cite{xu16} we likewise adopt the results of \cite{calzetti94} and take 
their internal dust model, where the ionized gas and dust are co-spatial and uniformly mixed (see the Appendix 
for an exploration of the alternative geometrical assumption of a foreground screen). In the present case

\begin{equation} \label{eqn_model_C_lum}
L_{\rm obs}(\lambda) =  L_{\rm i}(\lambda) 
  \cdot \frac{1}{\tau_\lambda} \left( 1 - e^{-\tau_\lambda} \right)
\end{equation}

\noindent where $\tau_\lambda$ is the total dust attenuation arising from both absorption and scattering. 
We use the resolved gas distribution of the simulation to calculate $\tau_\lambda^a$, the contribution 
from absorption. The scattering contribution is then incorporated by deriving a total $\tau_\lambda$ as

\begin{equation}
  \begin{aligned}
    \tau_\lambda &= f^{\rm eff}( \tau_\lambda^a, h_\lambda, \omega_\lambda ) \\
                 &= \tau_\lambda^a \left[ h_\lambda (1-\omega_\lambda)^{1/2} + (1-h_\lambda)(1-\omega_\lambda) \right]
  \end{aligned}
\end{equation}

\noindent where $h_\lambda$ and $\omega_\lambda$ are taken from fits to the 
scattering anisotropy weight factor and albedo over $\lambda \in $ [100\,nm, 1000\,nm], 
respectively \citep{calzetti94}. For the absorption optical depth, we take

\begin{equation}
\tau_\lambda^a = \left( \frac{A_\lambda}{A_V} \right)_\odot 
  (1 + z)^\beta \,
  (Z_g / Z_\odot)^\gamma \,
  (N_{\rm H} / N_{\rm H,0}).
\end{equation}

\noindent Here, the first term is the solar neighborhood extinction curve which we take from \cite{cardelli89}.
Note that in the optical the differences between the empirical extinction curves of the SMC, LMC, and Milky Way 
are negligible. The second and third terms parameterize a redshift and metallicity dependent dust-to-gas 
ratio \citep[see also][]{mckinnon16}. We adopt $\beta = -0.5$ and a broken power-law $\gamma = 1.6$ ($\lambda > 200$\,nm) 
$\gamma = 1.35$ ($\lambda < 200$\,nm, outside the SDSS bands). We take $Z_\odot = 0.0127$ and for the normalization of 
the hydrogen column $N_{\rm H,0} = 2.1 \times 10^{21}$\,cm$^{-2}$. All parameters are taken unchanged from their 
literature values.

The quantities $Z_g$ and $N_{\rm H}$ are both computed independently for each star with the following procedure.
For each galaxy, we trace through the resolved gas distribution from twelve viewing angles, corresponding to the 
verticies of the $N_s$\,=\,1 Healpix sphere \citep{gorski05} oriented in simulation coordinates (i.e. randomly with 
respect to each galaxy). For each direction, an orthographic projection grid 
with a pixel scale of $1.0$\,kpc is created covering all the member star particles. All gas cells and star 
particles in the subhalo are z-sorted by distance from the view point, then gas cells are progressively 
deposited on to the grid front-to-back using the standard cubic spline kernel with 
$h = 2.5$\,$r_{\rm cell} = 2.5$\,$(3 V_{\rm cell}/4/\pi)^{1/3}$. When the z-position of a star is reached it 
samples with bilinear interpolation the accumulated gas values in the four grid cells nearest to its projected location.
These values are the neutral hydrogen column density and the neutral hydrogen mass weighted gas metallicity, 
which are used for $N_{\rm H}$ and $Z_g$ above. Note that the neutral hydrogen fraction of each gas cell is 
computed during the simulation using the atomic network of \cite{katz96} with the time variable though spatially 
uniform ionizing UV background of \cite{fg09} given the density-based self shielding prescription of 
\cite{rahmati13}.

Each full stellar population spectrum is convolved with the $\lambda$-dependent attenuation and then by the observed 
passband. Finally, we impose a 3D radial restriction of at most 30 physical kpc from the subhalo center. This 
approximates a Petrosian aperture, and includes all subhalo particles for low-mass galaxies.


\section{Comparison to Observations} \label{sec_obs_comp}

To assess the outcome of the simulations we compare to the observed colors of galaxy populations in the local 
universe. Specifically, we draw an observed sample from the Sloan Digital Sky Survey \citep[SDSS DR12;][]{sdss_dr12} 
by selecting all galaxies with spectroscopic redshifts $z$\,$<$\,0.1, a sample of 378,287 objects. For their 
absolute magnitudes in the five \textit{ugriz} SDSS broadband filters we adopt the 
modelMag values. These are computed from the best fit of either an exponential or de Vaucoulers profile in the 
r-band, whose amplitude (but not shape) is then fit in the remaining bands.
They are in excellent agreement with Petrosian magnitudes for the majority of galaxies, with offsets due to color 
gradients and aperture corrections at the massive end \citep{strauss02}. We add the correction for Galactic extinction 
following \cite{schlegel98} as well as a k-correction using the method of \cite{chilingarian10}. For stellar mass 
estimates for each observed galaxy we adopt the Granada FSPS wide SED fit values \citep{ahn14}.

\begin{figure*}
\centerline{
\includegraphics[angle=0,width=3.5in]{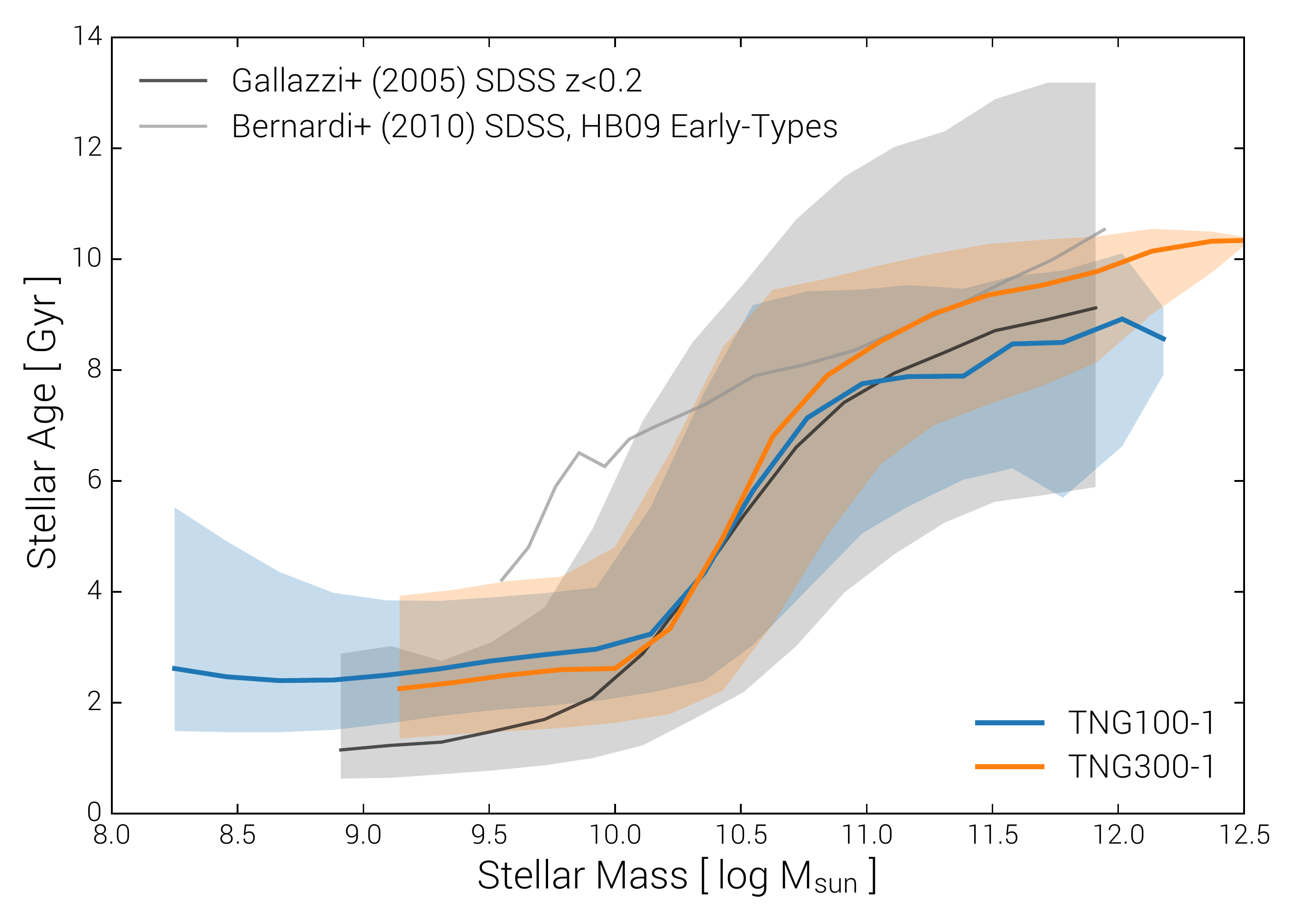}
\includegraphics[angle=0,width=3.5in]{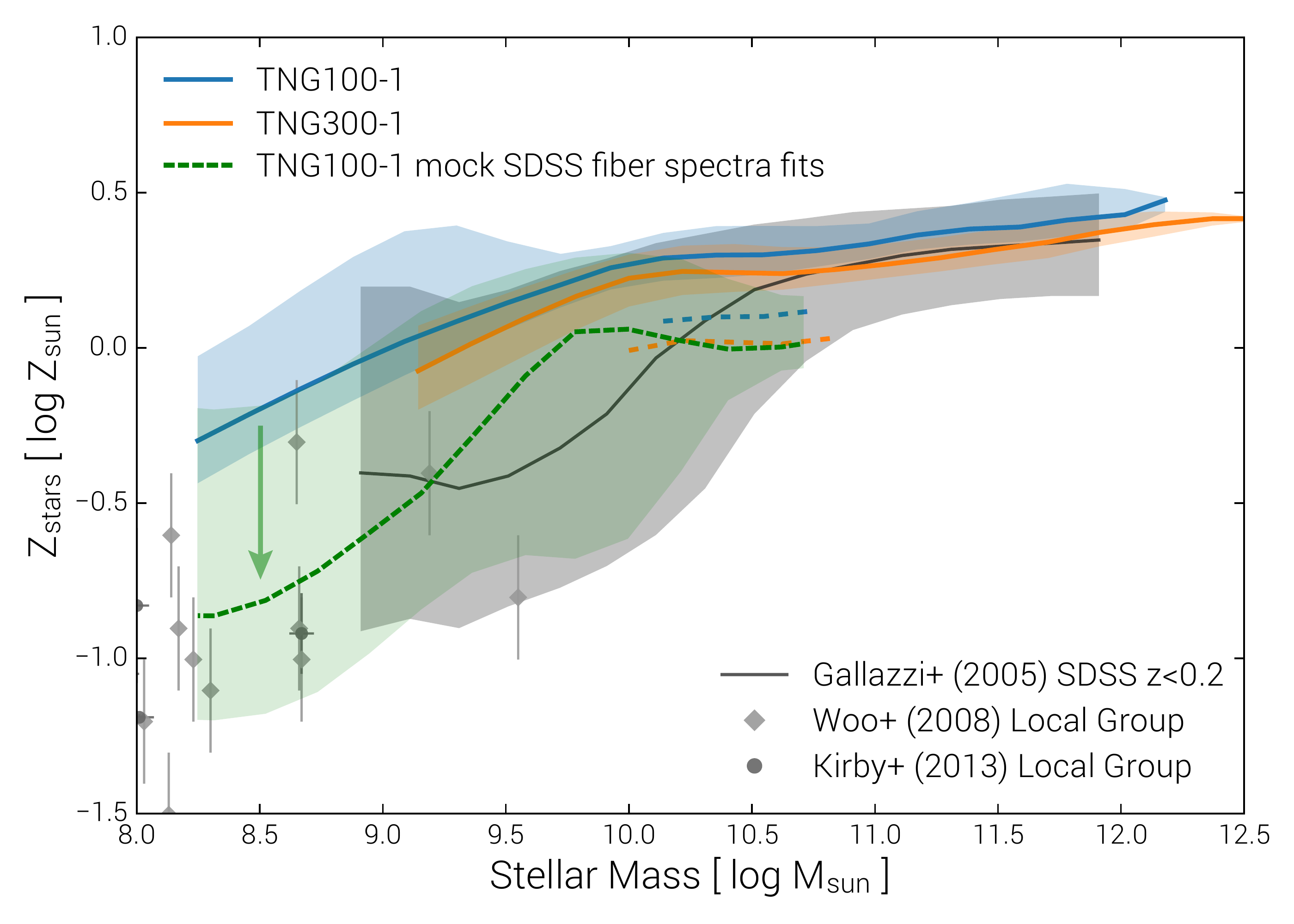}}
\caption{ The simulated stellar age versus mass (left panel) and stellar metallicity versus mass (right panel) relations, compared 
to several observational data sets (gray). The ages are compared to observational results from \protect\cite{gallazzi05} and 
\protect\cite{bernardi10}, while the stellar metallicity relation is compared to data from \protect\cite{gallazzi05} with dwarf 
points from \protect\cite{woo08} and \protect\cite{kirby13}. In both cases, the simulated lines are taken at $z\,=\,0.1$ by considering 
the mean age and/or metallicity of all star particles within a 4\,kpc mock SDSS fiber aperture, weighted by their r-band luminosities. 
For $Z_{\rm stars}$ we include approximate corrections from \protect\cite{guidi16} for the difference between our luminosity weighted 
fiber-aperture technique and the spectral index fitting technique actually used in \protect\cite{gallazzi05} (dashed blue and orange 
lines). In green we show the TNG100-1 trend and its scatter derived by fitting mock SDSS spectra with matched noise and resolution 
characteristics (see footnote for details).
 \label{fig_stellar_ages_Z}} 
\end{figure*}

In Figure \ref{fig_colors_1d_histos} we show normalized PDFs of (g-r) color in six equal bins of stellar mass from 
$10^{9}$ to $10^{12}$\,\msun\!. The black curves in each panel show the observed distribution of galaxy colors in that 
mass range. In red we show the results of the original Illustris simulation, while in blue we show the outcome 
of the new TNG100 simulation. We adopt the same dust model (`resolved dust' model C) for both, and include all 
$z\,=\,0$ galaxies with $10^{9}<M_\star/$\msun$<10^{12}$.
In both the observations and in TNG100 there is a consistent overall picture: at low stellar masses, the blue 
population dominates the sample with \mbox{(g-r)$\,\sim\,$0.4}, while the red peak emerges towards higher $M_\star$ at a 
characteristic \mbox{(g-r)$\,\sim\,$0.8}. The two reach roughly equal magnitude, displaying the strong observed color 
bimodality, at $M_\star \simeq 10^{10.5}$\,\msun\!. Above this mass blue systems quickly disappear, leaving the red 
galaxy population dominant at high mass.

The agreement between TNG100 simulated galaxies and the observational dataset is in general excellent -- see Section \ref{secCMPlane}. For 
stellar masses below $10^{11}$\,\msun where there are clearly two distinct peaks, we recover their respective locations 
in (g-r) and the position of the minimum between them to $\simeq 0.05$ mag or better. The evolution of the relative 
amplitude of the blue and red distributions as a function of $M_\star$ is likewise well captured, as is the stellar mass 
of maximal bimodality, where the red fraction of the galaxy population is half. On the other hand, the original 
Illustris simulation does not exhibit a color bimodality sufficiently strong, or at the correct location in $M_\star$, 
having generally too few red systems at intermediate masses between $10^{9.5}<M_\star/$\msun$<10^{11}$. Above 
$10^{11}$\,\msun Illustris agrees well with SDSS in the peak location of the red sequence, although the distribution may 
be too narrow in color at the highest masses.

The clearest potential tension between TNG100 and the $z$\,$<$\,0.1 SDSS sample is that the simulation appears to 
contain too many red galaxies at the lowest masses, below $10^{9.5}$\,\msun\!. In this panel (upper left) we also 
show the separate color distribution for central galaxies only, which demonstrates that this overly populated red 
component is entirely made up of satellites. TNG100 may genuinely produce too many low-mass red satellite 
galaxies. These are however some of the most marginally resolved systems, containing only $\sim$1000 star particles, 
and in Appendix \ref{sec_appendix} we show that with increasing resolution the low-mass galaxy population tends to 
become bluer. This may be, therefore, an unresolved regime for TNG100. On the other hand, the observations are not 
complete at these low masses. A bias in the effective volume sampled by the SDSS spectroscopic survey as a function 
of color, such that intrinsically fainter red galaxies are underrepresented, could explain the discrepancy. 
Similarly, a bias from the spectroscopic targeting due to fiber collisions in high-density environments, such that 
intrinsically redder satellites are underrepresented, could also contribute. A more sophisticated analysis -- namely,
a full mock SDSS spectral survey -- could definitely demonstrate the impact of both issues.

\begin{figure*}
\centerline{\includegraphics[angle=0,width=7.0in]{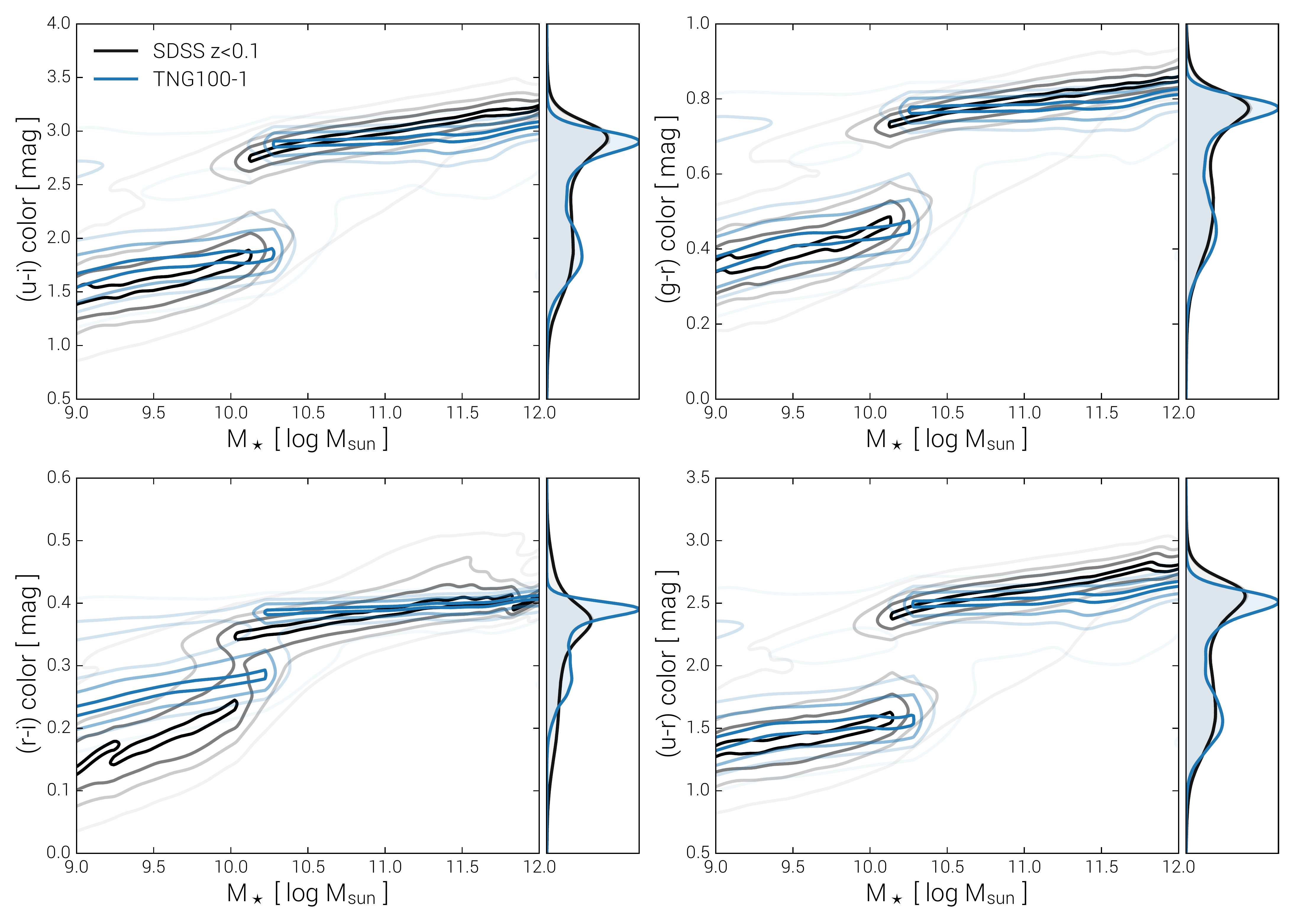}}
\caption{ The distribution of observed (black) versus simulated (blue) galaxies in the color-mass plane at $z$\,=\,0, 
shown for three band combinations beyond the (g-r) explored in this work: (u-i), (u-r), and (r-i). Conditional 2D 
kernel density estimates (KDEs) are shown with contours at \{0.2,0.5,0.75,0.98\}. On the right side of each panel 
the marginalized one-dimensional PDF is shown, drawing a simulated sample with the same stellar mass distribution 
as the observations using a discrete inverse transform sampling method. We see that the (u-i) and (u-r) colors show 
the same level of excellent quantitative agreement with respect to the observations, while the (r-i) colors of low 
mass galaxies are slightly too red.
 \label{fig_color_mass_2d_4bands}} 
\end{figure*}

\subsection{Stellar Ages and Metallicities}

The principal ingredients of the colors are the ages and metallicities of the stars. In Figure \ref{fig_stellar_ages_Z} 
we therefore independently compare the predictions for TNG100 and TNG300 at $z\,=\,0.1$ against observed stellar 
age and metallicity relations. The primary constraint for both is from \cite{gallazzi05}, which is also based on 
low-redshift SDSS galaxies. From the simulations we compute luminosity weighted r-band 2D fiber-aperture 4\,kpc 
restricted values as an approximation of the SDSS fiber derived observations.

The simulated stellar ages are in excellent agreement with the observed relation, both in the median values (solid lines) 
and in the scatter (16th-84th percentiles). We caution that the simulated scatter is intrinsic only -- we have not 
convolved by an estimate of the observational uncertainty. The simulated scatter is therefore consistent with the observed 
scatter in the sense of a lower limit. The average stellar ages transition from a relatively constant $\simeq 8-10$\,Gyr 
at the highest masses to a relatively constant $\simeq 2$\,Gyr at the lowest masses, with intermediate values over 
$10^{10}<M_\star/$\msun$<10^{11}$. This same trend, and the transition point at Milky Way mass, is seen in the observed 
ages. Tension exists in the value of the mean age of stars in the lowest mass galaxies, below $10^{9.5}$\,\msun\!. 
While \cite{gallazzi05} favor an asymptotic value of 1 Gyr, we find a somewhat higher mean age of 2 Gyrs. This lack 
of present day low-mass dwarf systems formed within the past billion years in the simulation may be a genuine 
failure of the model. As we discuss next, it may also arise from an overly simplistic comparison.

The stellar metallicities do not reproduce the observed trend to the same degree. At $10^{10.5}$\,\msun\! and above the 
simulations agree with the observed population quite well, exhibiting a weak scaling with $M_\star$ normalized at a super-solar 
metallicity value. However, the trend with stellar mass is too flat, and this leads to too high metallicities at low 
mass. Specifically, the maximal discrepancy is $\simeq 0.5$\,dex in stellar metallicity at $\log(Z_\star / Z_\odot) \simeq -0.5$ 
(observed), at the level of the 1$\sigma$ (upper 84th percentile) range of the observed sample. We hypothesize that 
this discrepancy is driven by the rather different method for deriving the simulated versus observed stellar metallicities. 
Specifically, we show (orange and blue dashed lines) a proposed correction between the two from \cite{guidi16}, which involves a dust 
radiative transfer computation of the emergent spectrum, followed by measurement of individual spectral features (e.g. D4000$n$, 
H$\beta$, [Mg$_2$Fe]) and the 
application of the specific procedure of \cite{gallazzi05}. This is rather distant from the simple luminosity-weighted 
fiber-aperture values we present, which do not involve any spectral measurements. Although we would like to use this 
correction directly, it has been derived from a small simulated sample over a limited mass 
range, and with a different physical model. A robust comparison of TNG stellar metallicities (and ages) to SDSS fiber 
spectrally-derived values therefore requires more sophisticated forward modeling.\footnote{A preliminary analysis -- whereby 
we have created mock SDSS fiber spectra for every simulated galaxy with noise and instrumental resolution characteristics 
sampled randomly from real SDSS fibers, identically performed a MCMC fit on both the mock spectral catalog and the actual 
SDSS spectral sample, using FSPS as the generative model \citep[][Johnson et al. in prep]{leja17} with the MILES libraries 
over 3750\AA$\,<\lambda<\,$7000\AA\, in the seven-dimensional parameter space of $\{z_{\rm residual}$, $M_\star$, 
$Z_\star$, $\tau_{\rm SFH}$, $t_{\rm age}$, $\tau_{\rm 1,dust}$, $\sigma_{\rm disp}\}$ with broad priors on each 
-- suggests that the correspondingly derived \textit{median} stellar metallicities are offset lower by $\sim 0.2$ dex 
at $M_\star = 10^{10}$\,\msun\! and by $\sim 0.5$ dex at $M_\star = 10^{9}$\,\msun\! with respect to the presented 
simulation values. This implies that the apparent tension in the $Z_\star-M_\star$ relation arises solely from an 
overly simplistic comparison (\textcolor{blue}{Nelson et al. in prep}). Note that we truncate this mock Z$_{\rm stars}$ 
measurement at $M_\star \simeq 10^{10.5}$\msun, after which our current modeling implies a similar and roughly constant 
offset extends up to the more massive galaxies. A more robust conclusion in this regime, however, would require an improved 
fitting of systems with significant stellar velocity dispersions.}

Indeed, the colors of stellar populations are sensitively dependent on both age and metallicity. Yet, we have 
demonstrated excellent quantitative agreement between simulated and observed galaxy colors down to, at least, 
stellar masses of $10^{9.5}$\,\msun. Therefore, the apparent tension in the stellar age and metallicity 
relations at these same masses suggests that the \textit{direct} comparison of age and metallicity is presently 
less informative than the more indirect comparison of color, because of the difficulty of making a 
rigorous comparison of the simulated results with SDSS fiber-derived quantities.

\subsection{Characterizing the Color-Mass Plane} \label{secCMPlane}

Color combinations with the u-band will span the 4000$\r{A}$ break at low redshift, making them more sensitive 
indicators of quiescence and star formation history \citep{bruzual83}. We therefore augment the (g-r) 
colors presented thus far with several additional comparisons against the SDSS sample.
In Figure \ref{fig_color_mass_2d_4bands} we show contours of the full two dimensional distributions in the 
color-mass plane for four band combinations: \mbox{(u-i)}, \mbox{(g-r)}, \mbox{(r-i)}, and \mbox{(u-r)}, comparing each 
to the same $z\,<\,0.1$ SDSS galaxies. We see that each of \mbox{(u-r)}, \mbox{(u-i)}, and \mbox{(r-i)} share the same level 
of quantitative agreement with the observations as (g-r). The blue population in \mbox{(r-i)} color is excessively red by 
roughly 0.1 dex, and since \mbox{(r-i)} is sensitive to dust reddening this discrepancy may indicate tension in our dust 
modeling.

More interestingly, we see the emergence of several second order discrepancies. These would have been impossible to observe, 
much less characterize, without this excellent base level of agreement. Two are immediately obvious.

The first is the fact that the sharp transition from blue to red occurs at slightly too high $M_\star$ in the 
simulations, by about 0.1 dex. The immediate conclusion would be that galaxies transition to a red-dominated population 
slightly too late. We caution, though, that we have not modeled mock SDSS stellar masses of galaxies at the same level 
of rigor as their colors \citep[see][for work in this direction]{torrey15,guidi15,bottrell17}; 
0.1 dex is likely also less than the observational uncertainty in $M_\star$.

The second discrepancy is that the slope of the red sequence is flatter in the simulations than in the observations. 
This implies that there may be some physical effect tilting the observed color-mass relation at the high-mass end, which TNG 
either misses or captures less strongly. The physical origin is likely the residual trends of age and $Z_\star$ 
within the passive population \citep{loubser09}. It is possible that in TNG the mean stellar age and $Z_\star$ do not 
increase as strongly with stellar mass as they should. On the other hand, we find that the integrated colors of the massive 
red population are quite sensitive to aperture. We propose that galactic color gradients act to wash out this 
intrapopulation effect: because passive galaxies have measurable radial gradients in stellar age, metallicity, and color 
\citep{gonzalez15,cook16} integrated colors are sensitive to the radius within which they are measured, and our fixed 
30\,kpc aperture restriction cannot precisely correspond to the SDSS model magnitudes. Smaller apertures exclude 
a progressively larger fraction of (generally bluer) ICL, reddening the integral color of the simulated galaxy.
Similar considerations may explain why simulated galaxies in the most massive bin of Figure \ref{fig_colors_1d_histos} 
(at M$_\star \ge 10^{11.5}$ \msun) do not reach the maximal (g-r) values seen in SDSS, although the lack of this tail at 
(g-r) $>$ 0.9 may also reflect the inability of our dust modeling to represent the most extremely reddened observed galaxies.

\begin{figure}
\centerline{
\includegraphics[angle=0,width=3.3in]{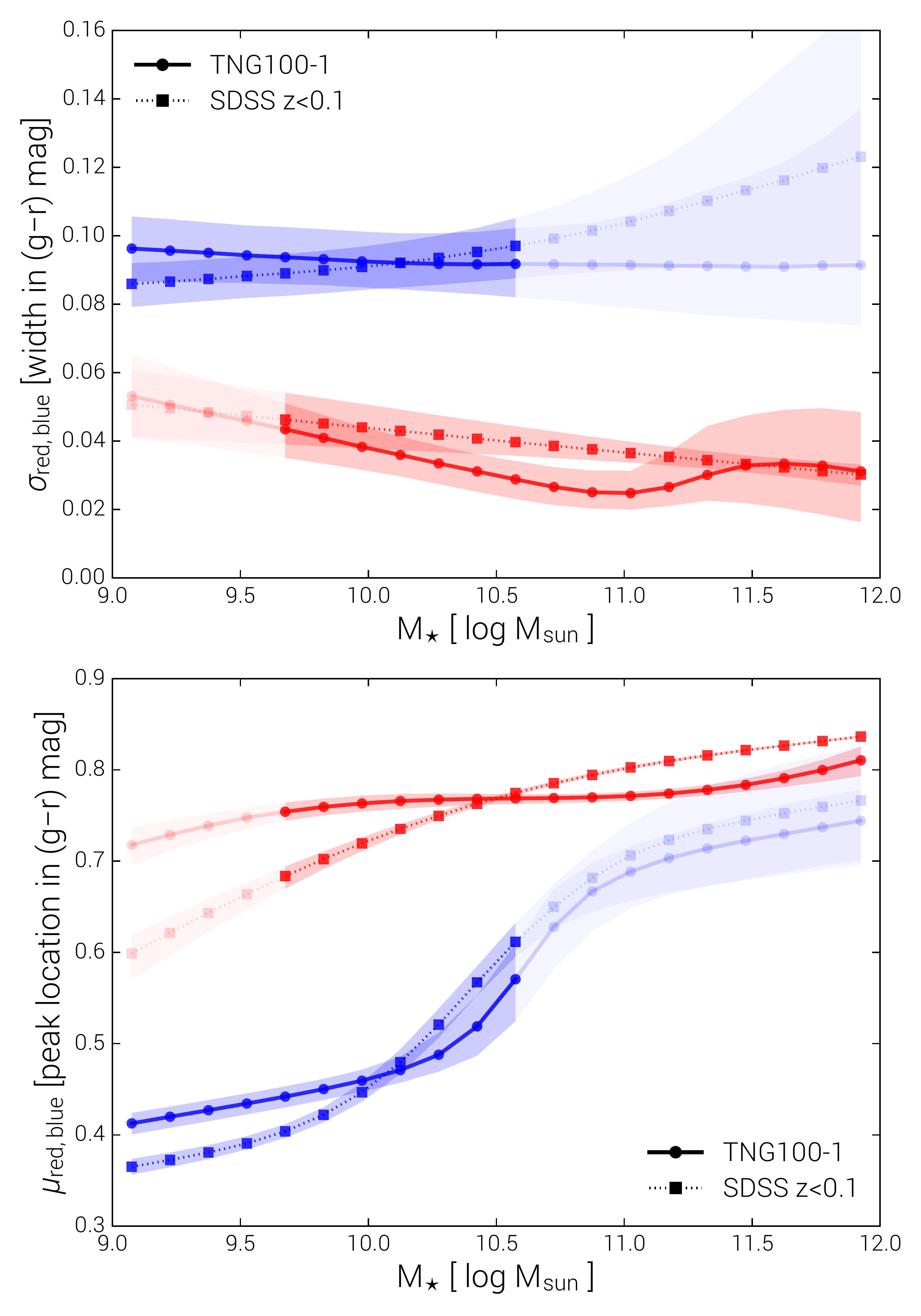}}
\caption{ Comparison of the red and blue population fit parameters: widths $\sigma_i$ (top panel) and center locations $\mu_i$ 
(bottom panel). The same procedure has been applied to the SDSS $z\,<\,0.1$ sample and the TNG100-1 simulation at $z\,=\,0$. 
We visually focus on the low-mass blue and high-mass red parameters by showing them as dark lines and shaded regions, which give 
the median and 10-90 percentile model ranges of the final chains, respectively.
 \label{fig_cm_plane_fits}} 
\end{figure}

To avoid ambiguous statements about having `reasonable', `good', or `excellent' agreement between observed and simulated 
galaxy properties we quantify the respective color distributions. Inspired by \cite{baldry04} and \cite{taylor15} we proceed 
without making a priori cuts on where red versus blue galaxies populate the color-mass space. Instead we assume that, at a given 
stellar mass, two distinct (possibly overlapping) populations exist, each with a gaussian distribution $\mathcal{G}(C)$ in 
color $C$. In a discrete set of mass bins, we therefore fit a series of 2-gaussian mixture models

\begin{equation}
\sum_{i=0}^{1} \mathcal{G}_i =
\sum_{i=0}^{1} \frac{ i + (-1)^i A_i }{\sigma_i (2\pi)^{-1/2} } \exp{ \left[ \frac{ -(C - \mu_i)^2 }{ 2 \sigma_i^2 } \right] }.
\end{equation}


\noindent We consider the stellar mass range spanning 9.0\,$<$ log(\,$M_\star$/\msun)$<$12.0 with a binsize $\Delta \log M_\star$ = 0.15 
giving 20 total mass bins. Similarly, a color binsize of $\Delta (g-r)$ = 0.04 over the color range 0.0\,$<$\,(g-r)\,$<$\,1.0 
gives 25 color bins per mass bin. Rather than let the five parameters of each $\sum \mathcal{G}_i$ vary independently, we 
constrain their center positions and widths to be continuous functions of stellar mass according to a global linear+tanh relation, namely

\begin{equation}
\mathcal{T}(M_\star) = p_0 + p_1 \log M_\star + q_0 \tanh \left[ \frac{\log M_\star - q_1}{q_2} \right].
\end{equation}

\noindent We have 20 relative amplitudes for the red versus blue populations across the mass bins. In addition, ten $\mathcal{T}(x)$ 
function parameters for the red and blue $\mu_i$, and likewise the ten $\mathcal{T}(x)$ function parameters for the red 
and blue $\sigma_i$, resulting in a forty parameter vector $\Theta$ to fit. Rather than a simultaneous global fit, we first 
implement and carry out the parameter-split iterative procedure as described in \cite{baldry04}. We then take this 
solution $\Theta_0$ as the initial guess of a global MCMC fit to explore the 40-dimensional 
parameter space and estimate the uncertainties of each \citep[using \textsc{emcee};][]{dfm13}. 

Applying the same procedure to the simulated and observed data sets, Figure \ref{fig_cm_plane_fits} shows the result. 
In it we compare the locations and widths of the red and blue populations as a function of stellar mass, in SDSS (dotted) 
versus TNG100-1 at $z\,=\,0$ (solid). The mean color of blue galaxies is in excellent agreement, within 0.05 mag. Its location in \mbox{(g-r)} 
may increase more slowly with $M_\star$ in the observations, although we caution that this could be driven 
largely by the balance between the $p_i$ and $q_i$ components of the $\mathcal{T}$ function (i.e., the amount of curvature is 
not well constrained). Most interestingly, we can quantify the red sequence discrepancies noted above: in the simulation the mean 
of this is slightly too blue (by $\la 0.05$\,mag). This tension arises because the dependence of $\sigma_{\rm red}$ on 
$M_\star$ is steeper than observed, while the dependence of $\mu_{\rm red}$ on $M_\star$ is shallower than observed. The simulated 
color widths are slightly less than observed, indicating that the intrinsic scatter is consistent with the observed scatter. We 
reiterate that we have not convolved the intrinsic colors with estimated observational uncertainties, which would be required to 
compare the scatter in detail. In addition, we here use (as always) the fiducial model (C) `resolved dust' colors for the 
simulations, and note that the exact level of quantitative (dis)agreement with respect to SDSS must depend at some level on the details 
of our dust modeling assumptions, which we explore in the Appendix.


\begin{figure*}
\centerline{\includegraphics[angle=0,width=7.0in]{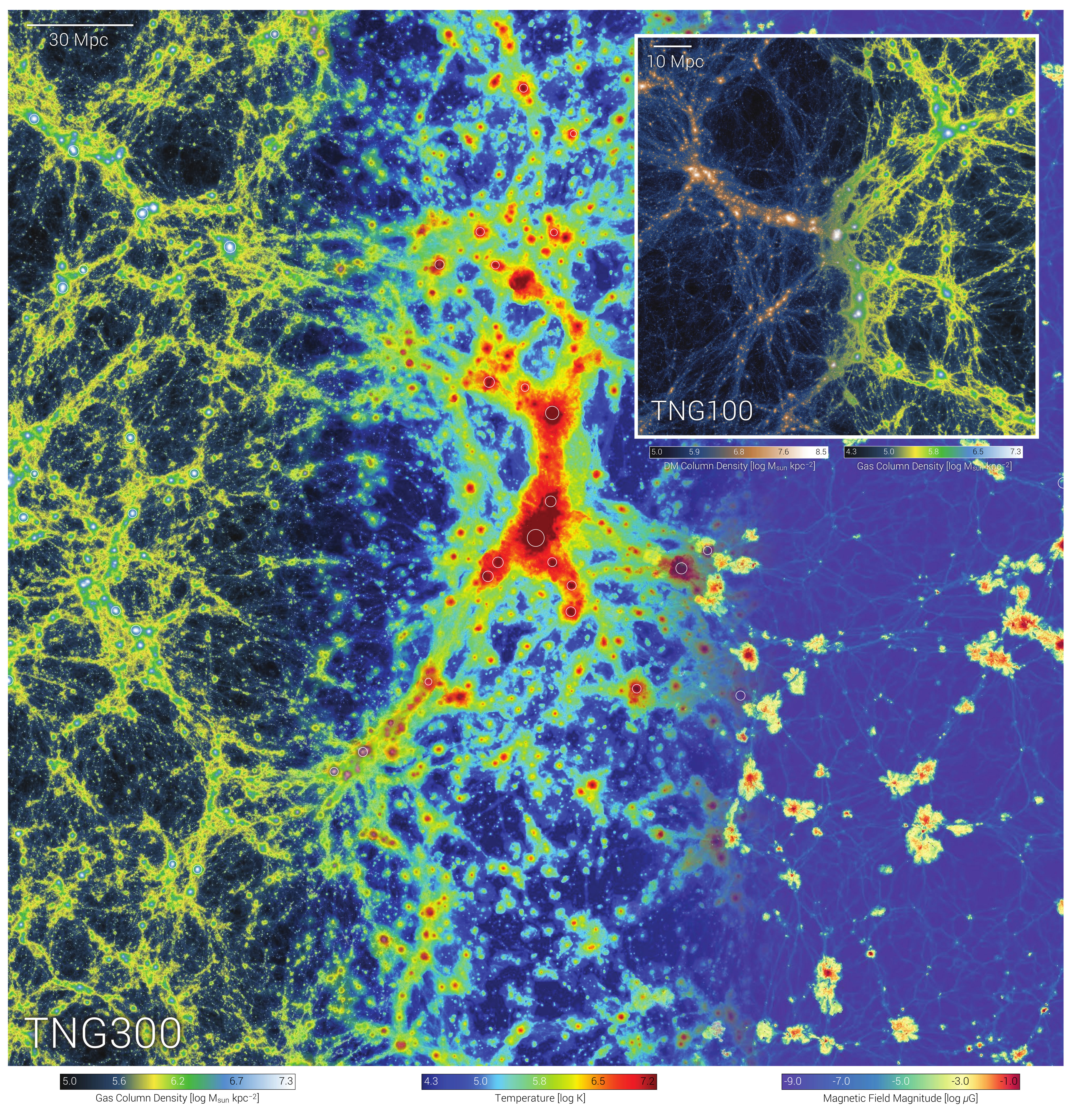}}
\caption{ The cosmological context offered by the TNG100 and TNG300 simulations. The main figure shows three views of TNG300-1 at 
$z\,=\,0$, centered on the most massive cluster in the box: gas column density (left), gas temperature (center), and magnetic field 
amplitude (right). The first two are projected through a depth equal to a third of the volume, while the magnetic field is shown in a 
thin 100 kpc slice. The 50 most massive halos in the thick slice are indicated by circles with radii of $r_{\rm 200,crit}$. 
In the inset in the upper right, we show the TNG100-1 volume at $z\,=\,0$ in projected dark matter density (left) and gas density (right). 
 \label{fig_fullbox_demo}} 
\end{figure*}

\section{Theoretical Interpretation} \label{sec_theory_interp}

\begin{figure*}
\centerline{\includegraphics[angle=0,width=7.0in]{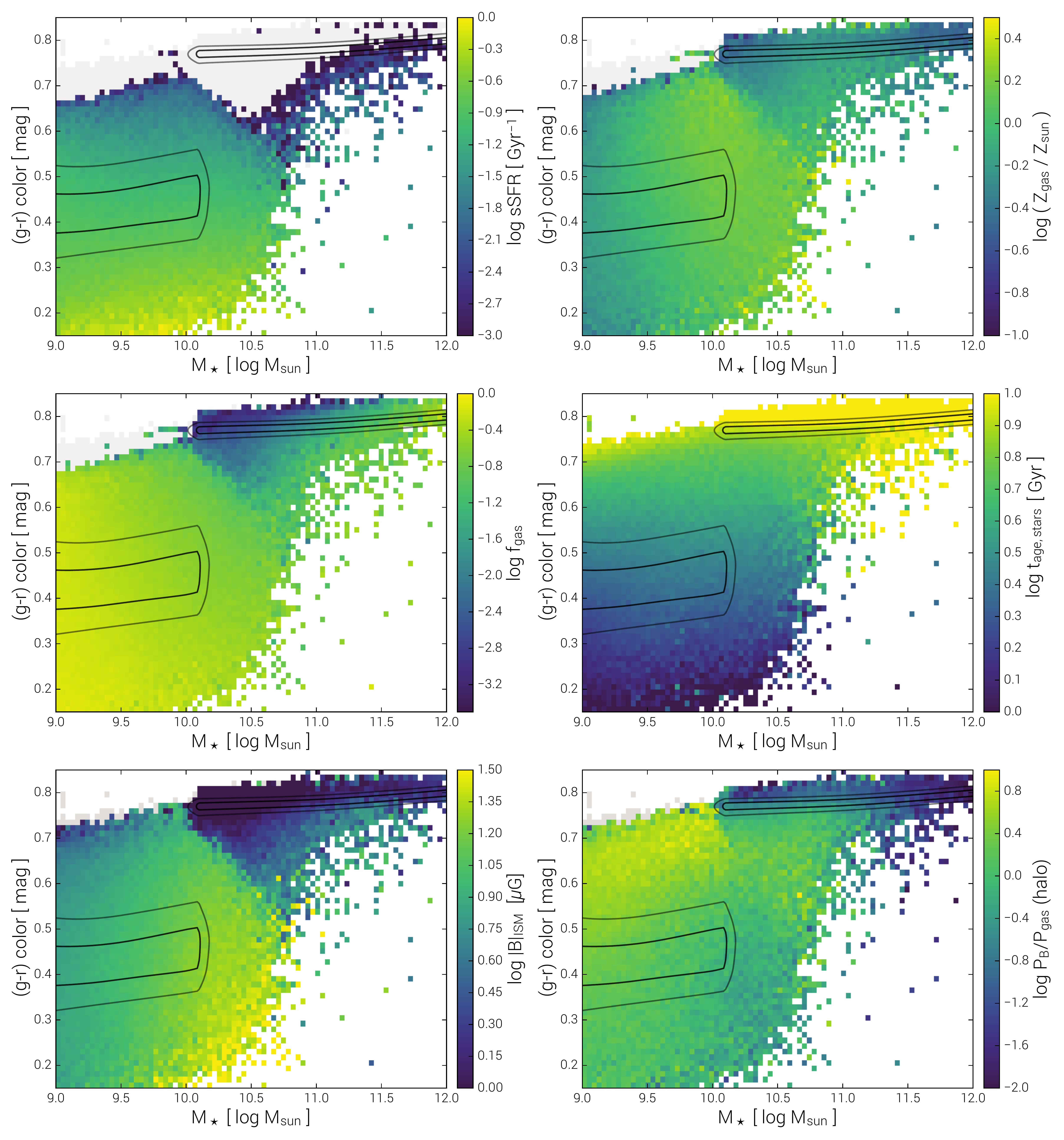}}
\caption{ Six views of the color-mass plane of all central galaxies in TNG300-1 at $z$\,=\,0. In each case colors show a physical, 
median value of all the galaxies in that bin, and measurements are always within $2r_{\star,1/2}$ unless otherwise noted. We show
(i) the specific star formation rate, 
(ii) the gas metallicity normalized by solar,
(iii) the gas fraction -- gas mass divided by total baryon mass,
(iv) the mean r-band luminosity weighted stellar age within a 4\,kpc fiber-aperture,
(v) the mean magnetic field strength in the ISM, and
(vi) the $\beta^{-1}$ parameter of the mean ratio of magnetic to thermal pressure in the \textit{halo} ($0.15 < r/r_{\rm vir} < 1.0$).
Gray indicates a zero (or n/a) value for a given property of all the galaxies in that bin. Contours show conditional 75\% and 95\% 
levels.
 \label{fig_2d_color_histograms}} 
\end{figure*}

Together, the TNG100 and TNG300 simulations are a powerful theoretical tool. In Figure \ref{fig_fullbox_demo} we show a 
visual overview of the two simulation volumes at the present day, $z\,=\,0$, including several properties of the gas as well as the 
dark matter distributions. The light of stars is entirely absent from this composite view, while all the galaxies (together with 
their colors) which we have thus far studied exist within this full cosmological context.
By construction these large volumes include galaxies of a great variety of types, environments, and formation histories. Simultaneously, 
each system has a large information content, whereby a wide range of properties of its stars, gas, dark matter, or blackholes 
can be measured. All these components have arisen self-consistently as the result of a coupled physical evolution across the history of 
the universe. In this section we use these relationships to elucidate the origin of the galaxy color distribution. We frequently include 
the TNG300 large volume in order to take advantage of its superior statistics, and refer to the Appendix for a discussion of resolution 
convergence and the differences between TNG300 and the higher resolution TNG100 simulation.

\subsection{Relating Color to Other Galaxy Properties}

In Figure \ref{fig_2d_color_histograms} we show six views of the same TNG300-1 color-mass plane at $z\,=\,0$. In each case, we map 
color to one median property of all the galaxies falling in that specific bin. These are: the specific star formation rate 
(upper left), gas metallicity (upper right), galaxy gas fraction (center left), galaxy stellar age (center right), ISM magnetic field 
amplitude (lower left), and \textit{halo} magnetic to gas pressure ratio (lower right). Gray colors indicate a uniformly zero value 
for that bin, while contours show the conditional 75\% and 95\% levels of the galaxy population as a function of $M_\star$.

In broad comparison of the lower left to the upper right of each panel -- the locations of the blue and red populations, respectively -- 
we recover several expected results. Red galaxies are passive, gas poor, and old. With respect to otherwise increasing trends with 
stellar mass, they also have lower gas metallicities and smaller magnetic field strengths, the latter in both the galaxy and the 
halo. Stellar age is the only property shown which depends almost entirely on color and not at all on $M_\star$, indicated by the 
purely vertical gradient. The fact that more massive galaxies tend to be older therefore imprints itself as an increasing red fraction 
with stellar mass. On the other hand, galactic stellar metallicity has a largely horizontal gradient in this plane (not shown), 
reflecting the stronger dependence of color on age.

The other five properties we show here, all related to the gas, have more complex behavior. Each appears driven by an underlying 
scaling with galaxy mass, which is then strongly modulated by color. For instance, the mean galactic gas metallicity increases with 
$M_\star$ following the mass-metallicity relation only while the majority of galaxies are star-forming. Above a threshold stellar mass 
of $\sim$ 10$^{10.5}$ \msun we see that $Z_{\rm gas}$ is suppressed. Similar trends are clearly visible in $f_{\rm gas}$ as well as 
$|B|_{\rm ISM}$. The emergent sSFR lacks any significant mass dependence, but otherwise has a similar trend. The halo $\beta$-ratio 
is the most complex, with non-monotonic behavior along both axes which we shortly explore further.

In the transition regime of $M_\star \sim 10^{10.5}$\,\msun and at \textit{fixed} stellar mass, the statistics of TNG300 allow us to 
see by eye the correlations between galaxy color and all five of the gas properties. For instance, in the median specific star 
formation rate, a vertical slice at this mass scale transitions from highly star forming (in yellow) at a (g-r) color of 0.3 to 
complete quiescence (in purple/gray) at a (g-r) color of 0.7. Similarly for the gas fraction, which also drops from near unity to 
essentially zero at a fixed stellar mass of $\sim 10^{10.5}$\,\msun\!. We point out an explicit prediction of the TNG model: at the 
Milky Way mass scale blue galaxies should have a large-scale interstellar magnetic field strength of order 10-30 $\mu$G while 
red galaxies, at the same stellar mass, should host a much weaker field of $\la$\,1\,$\mu$G.

An intriguing feature evident in Figure \ref{fig_2d_color_histograms} is the diagonal boundary dividing the red and blue regimes. 
It implies the existence of a mass-dependent color threshold, or equivalently stated, a color-dependent mass threshold, 
beyond which galaxies enter a distinct physical population. We show later that this diagonal separation in the CM plane 
does not correspond to a boundary orthogonal to the bulk flux of galaxies as they evolve, which is more vertical. Still, 
its existence, location in $M_\star$, and slope are all signatures of the color transition mechanism, driven by the specifics 
of the quenching process of the TNG blackhole feedback model.

\begin{figure}
\centerline{\includegraphics[angle=0,width=3.4in]{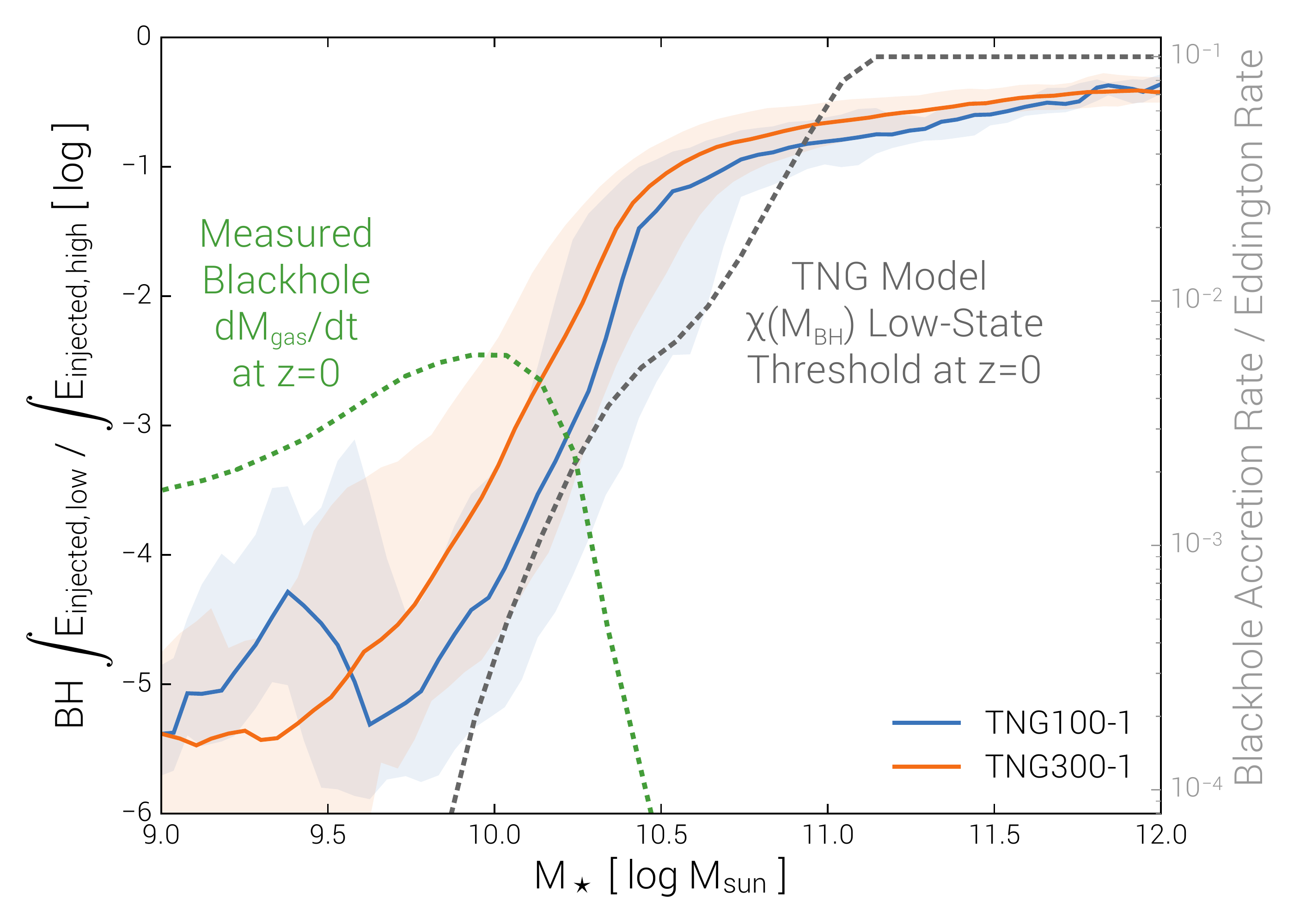}}
\caption{ The role of supermassive blackhole feedback in setting the characteristic stellar mass of color transition: the logarithm 
of the ratio of the integrated energy injection in the low versus high accretion states. These two quantities are summed across 
all blackhole progenitors. The median across all central galaxies is 
given for TNG100-1 (blue) and TNG300-1 (orange). This ratio is always less than one, increasing rapidly between 
\mbox{10.0 $<$ log($M_\star/$\msun\!) $<$ 10.5}, similar to the transition from a blue to red dominated galaxy population. 
In addition we show the value of $\chi(M_{\rm BH})$ defining the fraction of the Eddington accretion rate below which a blackhole is 
in the low state (dashed gray), and the actual $z=0$ blackhole accretion rates (dashed green), which intersect at this same characteristic 
mass.
 \label{fig_bh_cumegy}} 
\end{figure}

\begin{figure}
\centerline{\includegraphics[angle=0,width=3.3in]{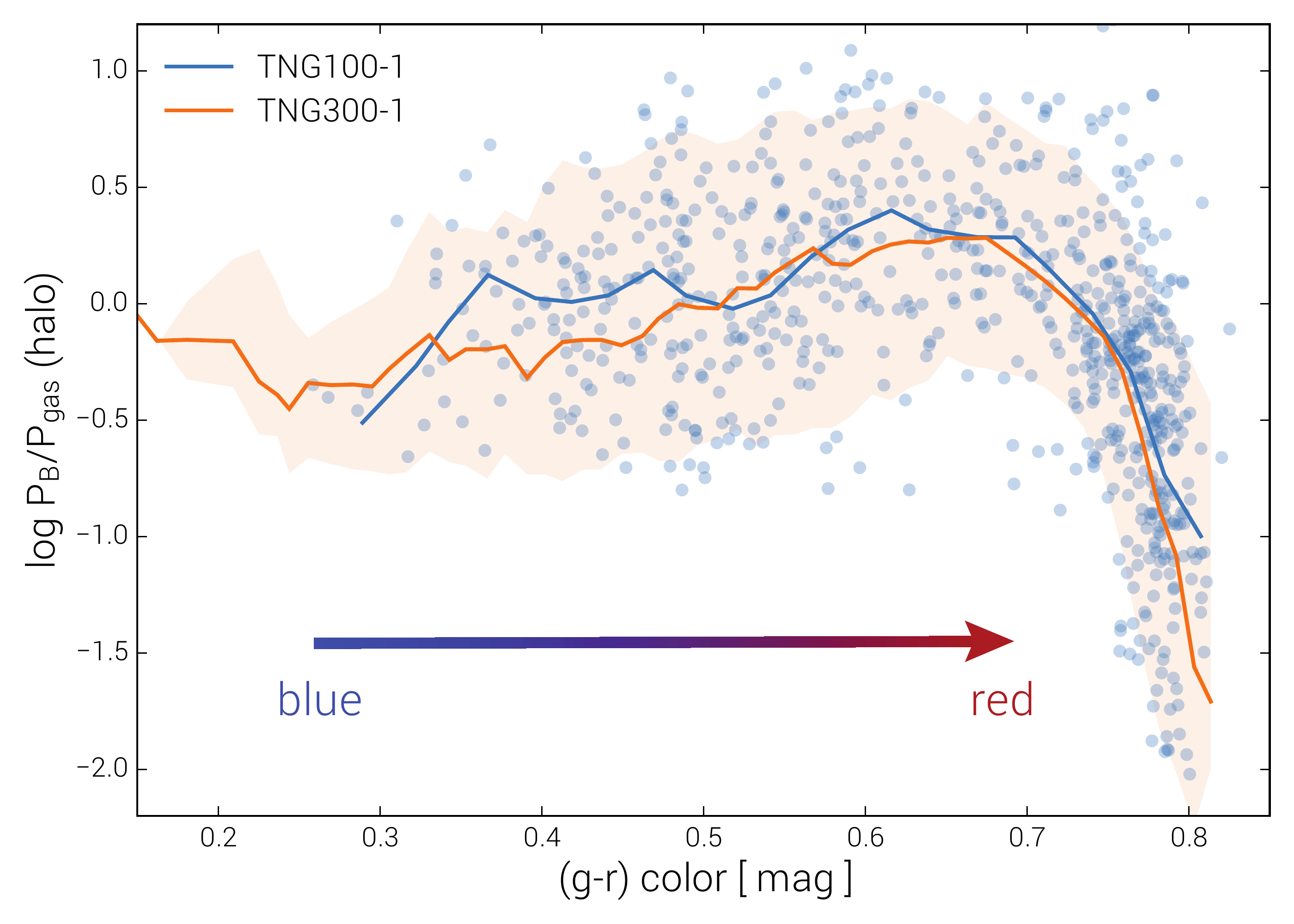}}
\caption{ The correlation between the $\beta^{-1} = P_{\rm B}/P_{\rm gas}$ pressure ratio of the halo gas and the (g-r) color of the 
galaxy at a stellar mass of $10^{10.5}$\,\msun\!. We include all central galaxies for TNG100-1 (blue median line and individual 
galaxies indicated by circles) as well as TNG300-1 (orange median line and 10 to 90 percentile shaded region) in a mass bin of 
width $\pm 0.1$\,dex. Blue galaxies have $\beta^{-1} \sim 1$, which then increases moderately with increasing color 
for transitional systems, before reversing and plummeting to $\beta^{-1} \sim 10^{-2}$ in the halos of the reddest galaxies. 
This trend is a \textit{signature} of the physical mechanism of color transformation.
 \label{fig_slice}} 
\end{figure}

To identify the underlying physical mechanism of color transition, Figure \ref{fig_bh_cumegy} shows a diagnostic of the 
integrated energy input from feedback by supermassive blackholes. Specifically, the cumulative energy released in the low accretion 
state (kinetic mode) is compared to the cumulative energy released in the high accretion state (quasar mode), as a function of 
$M_\star$ of the central galaxy. For low-mass galaxies this ratio is effectively zero as these SMBHs have never been in the 
low-state. For the most massive galaxies this ratio asymptotes to $\sim$\,0.5, but is always less than one. A rapid transition 
between these two regimes occurs between $10^{10}$\,\msun and $10^{10.5}$\,\msun, coinciding with the stellar mass where 
the median color of galaxies evolves from predominantly blue to predominantly red.

\begin{figure*}
\centerline{
\includegraphics[angle=0,width=3.5in]{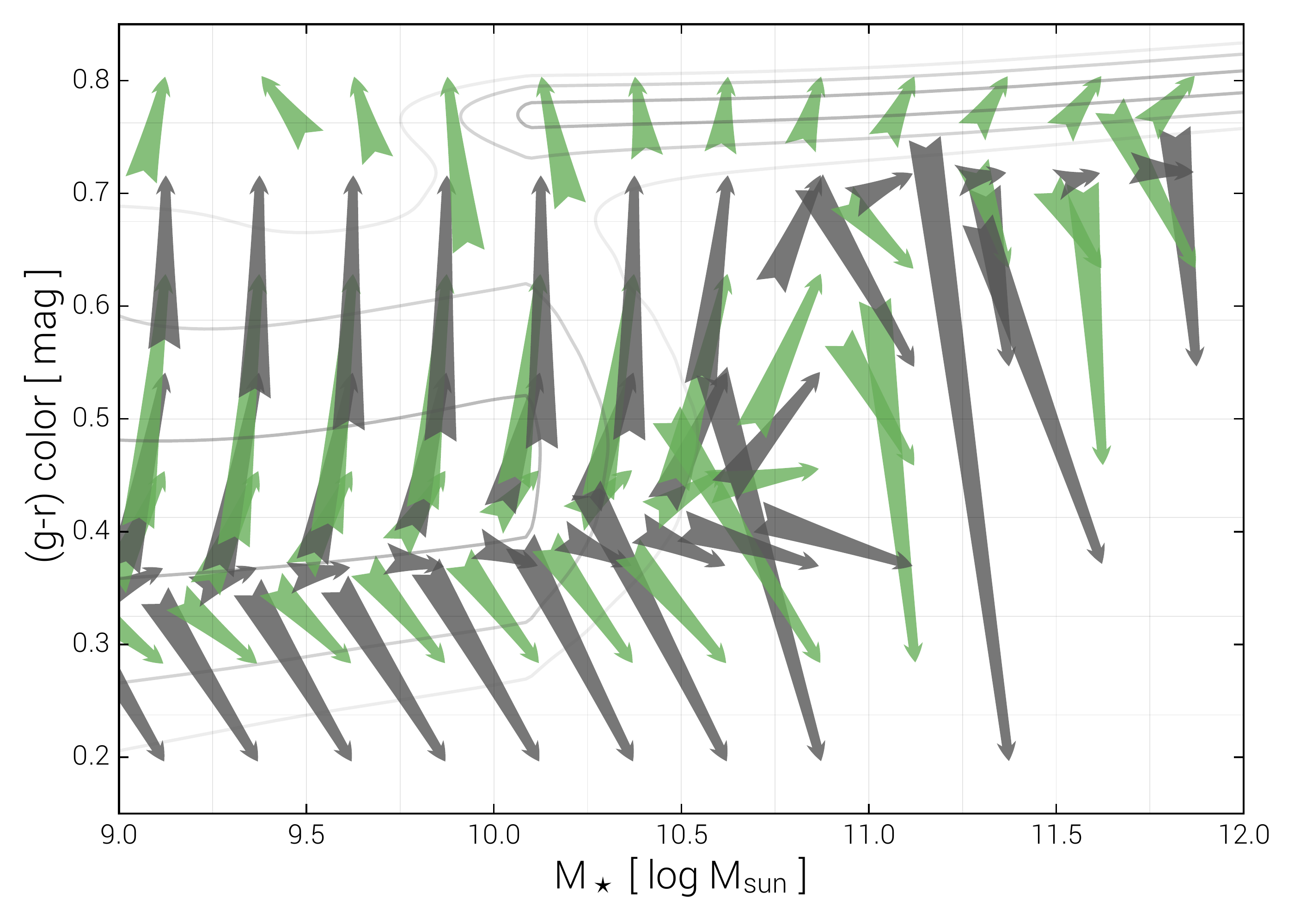}
\includegraphics[angle=0,width=3.55in]{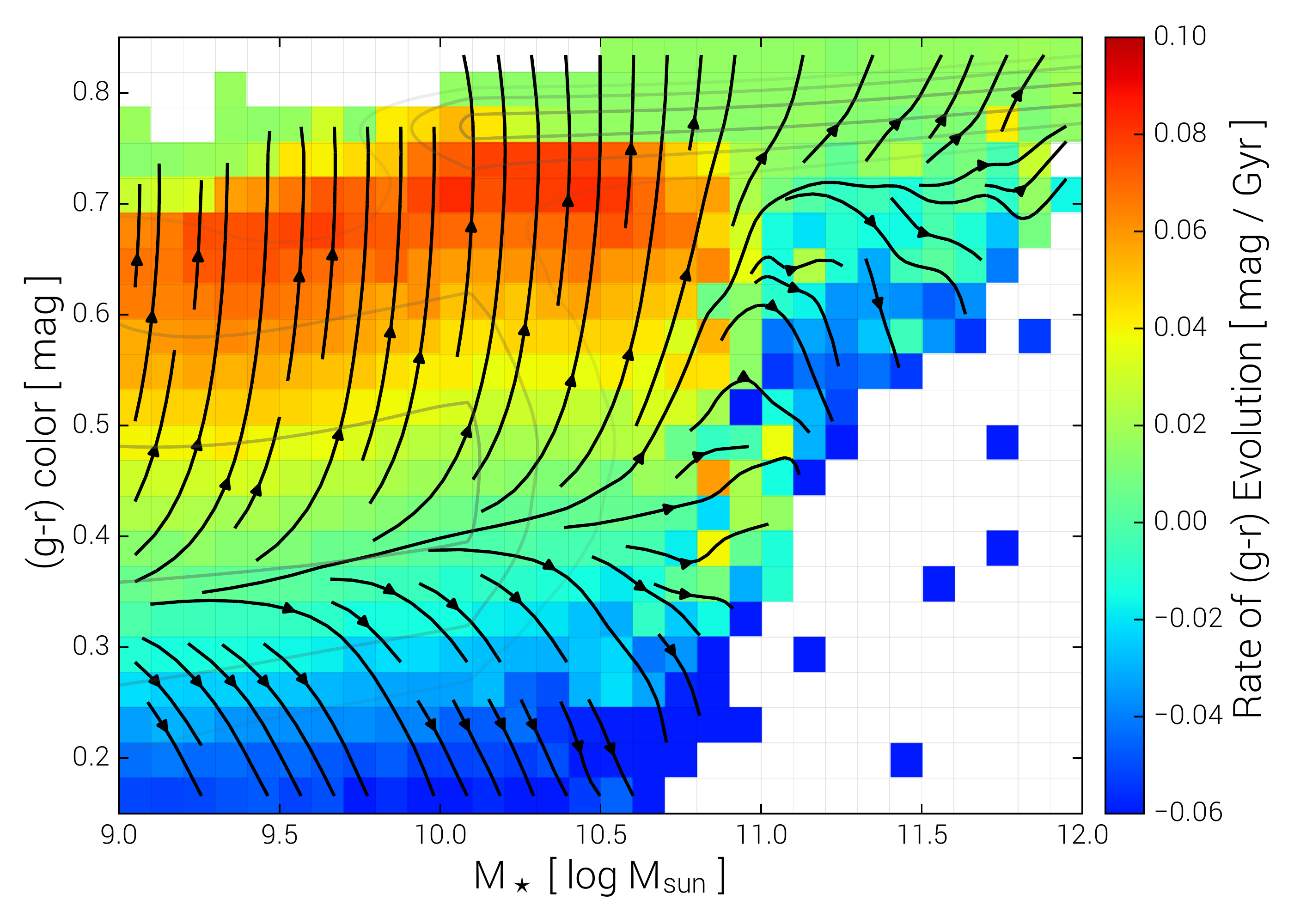}}
\caption{ The flux of galaxies across the color-mass plane with time in TNG300-1. Both panels include all redshift zero central galaxies 
and show evolution between $z\,=\,0$ and $z\,=\,0.3$, corresponding to 3.4\,Gyr -- the results are not sensitive to this time interval. 
Contours show the locations of the red and blue populations at redshift zero for 
reference. Left panel: individual arrows show the median origin $M_\star$ and (g-r) color of all the galaxies which arrive to the 
indicated bin by $z\,=\,0$. The background grid shows the binning, with green and gray arrows on alternative rows. Right panel: the 
background color measures the median rate of color evolution, computed from all the galaxies in that bin at redshift zero, in 
magnitudes per billion years. Overlaid are streamlines (in black) showing the direction of the local vector field 
$(\partial M_\star / \partial t, \,\partial (g-r) / \partial t)$ of galaxy evolution at that point. Several evolutionary phases 
are evident including (horizontal) mass growth along the blue population and (vertical) reddening towards quiescence.
These figures reveal the dynamic and evolving nature of the color-mass plane by showing the bulk evolution of galaxy ensembles, 
and do not necessarily represent the pathway of any individual system.
 \label{fig_cmplane_fluxes}} 
\end{figure*}

In the TNG model, blackholes enter the low-state when their accretion rate relative to Eddington drops below a threshold 
$\dot{M}/\dot{M}_{\rm edd} < \chi(M_{\rm BH})$ which depends on the blackhole mass \citep[][Eqn. 5]{weinberger17}. Using the 
median $M_{\rm BH} - M_\star$ relation we plot the effective threshold in the same figure (dashed gray line), as well as the 
actual $\dot{M}_{\rm BH}$ accretion rates (dashed green line). As the stellar mass of a galaxy grows above 
$\sim$\,10$^{10}$\,\msun BH accretion drops while the value of $\chi$ also rises, increasingly promoting transition into 
the low-state. Note that the model \textit{input} $\chi$ was set to address the stellar mass content of massive halos and the 
location of the knee in the stellar mass function \citep{pillepich17a}. We therefore conclude that the mass scale where 
galaxies undergo rapid color transition is set by SMBH feedback in the low-accretion state. Furthermore, we find that an evolution 
from blue to red color at this particular mass is unavoidably linked to the steep decline of the high-mass end of the low redshift 
stellar mass function. 

Finally, in Figure \ref{fig_slice} we return to the ratio $P_{\rm B}/P_{\rm gas}$ of magnetic to thermal gas pressure support in the 
\textit{halos} of galaxies. Selecting all galaxies in the narrow mass bin \mbox{$10^{10.4} <$ $M_\star$/\msun $< 10^{10.6}$} we 
measure the median relation as a function of (g-r) color, effectively at fixed stellar mass. While blue galaxies have a typical 
magnetic to gas pressure ratio of order unity in the halo gas, this value increases by a factor of a few at intermediate colors, 
before dropping sharply by a factor of 100 for the reddest objects. That is, the relative importance of the halo magnetic field 
with respect to the halo thermal energy is enhanced during the color transition phase and becomes negligible after. We generally 
expect amplification towards equipartition and $\beta = 1$, as seen in the blue population. On the other hand, with 
$P_{\rm B} \ll P_{\rm gas}$ the magnetic field in the halo is effectively controlled by the gas thermodynamics. This sharp decline 
arises because $P_{\rm B}$ drops much faster than $P_{\rm gas}$, coinciding with the changing density and thermal energy content 
of the hot halo. The properties of the circumgalactic gas around a galaxy therefore encode information about the energy release 
by the central blackhole, the cessation of star formation, and the color transition.


\subsection{The Buildup of the Color-Mass Plane}

To untangle the origin of the $z\,=\,0$ properties considered thus far, we begin to examine the time evolution of galaxies, 
focusing on their motion through the two-dimensional space of (g-r) color and stellar mass. In Figure \ref{fig_cmplane_fluxes} 
we show two views of the flux of galaxies across the color-mass plane from $z\,=\,0.3$ to redshift zero, a period of roughly 
3.4\,Gyr. In the left panel we show, for all the galaxies in each bin at \mbox{$z\,=\,0$}, where this ensemble of galaxies 
originated from. Arrows start at the median mass and color at \mbox{$z\,=\,0.3$} and end at the redshift zero location. In the 
right panel we show the rate of color evolution in the background color, superimposing streamlines derived from the vector 
field of $\Delta M_\star$ and \mbox{$\Delta$(g-r)} taken across this same time period. 

Along the lower edge of the $z\,=\,0$ blue population, (g-r) $\simeq$ 0.35, we see mass growth largely in the horizontal 
direction. From the right panel it is clear that this line acts as a separatrix in the population between either a positive 
or negative \textit{mean} change of galaxy (g-r) color with time. Below this line, the population becomes more blue as 
stellar mass increases, while above the opposite is true, leading to the butterfly pattern in color-mass space. This line of 
demarcation extends all the way to $M_\star = 10^{12}$\,\msun\!, although the origin at high mass differs. Above a stellar 
mass of $\ga 10^{11}$\,\msun it moves up to a (g-r) color of $\simeq$\,0.7, at the bottom edge of the red population. Here, 
galaxies above this line redden slowly due to passive stellar evolution, while those below undergo episodes of rejuvenated 
star formation which temporarily pollute their integrated colors with the light from young stars. From the left panel we 
see these galaxies can traverse the full color range, although the sparsity of high-mass blue systems at $z\,=\,0$ 
indicates these events are rare.

One feature not immediately evident is strong passive mass growth along the passive red population. The most massive 
galaxies must either grow in mass through a succession of dry mergers after becoming red at lower mass, or directly 
join the red sequence at essentially their final mass. 

\subsection{Timescale of the Color Transition and Formation of the Red Population}

To measure the evolution of individual galaxies in to, and out of, the red and blue populations, we first repeat 
the characterization of the color-mass plane as described in Section \ref{secCMPlane} over the range $0 \le z \le 1$. 
Colors are always derived from the rest-frame spectra of galaxies; changing colors indicate underlying physical evolution.
At each discrete redshift and in every stellar mass bin this yields independent fits of 
$\mu_{\rm red}, \mu_{\rm blue}, \sigma_{\rm red}, \sigma_{\rm blue}$. At each $M_\star$ we then fit for the redshift 
evolution of each of these four parameters assuming a powerlaw in $(1+z)$. For example, at $M_\star = 10^{11}$\,\msun (in the 
red population) we find $\mu_{\rm red} \propto (1+z)^{-0.3}$ and $\sigma_{\rm red} \propto (1+z)^{0.8}$ such that from $z\,=\,0$ 
to $z\,=\,1$ the location of the red peak drops by $\simeq$\,0.15\,mag in (g-r) and its width increases by almost a factor of two.
At $M_\star = 10^{10}$\,\msun (in the blue population) we find $\mu_{\rm blue} \propto (1+z)^{-0.6}$ and 
$\sigma_{\rm blue} \propto (1+z)^{-0.4}$ such that, over the same redshift range, the blue peak moves down by 
$\simeq$\,0.2\,mag in (g-r) while its width decreases by $\sim$\,30\%.

This quantitative characterization of the evolving blue/red galaxy distributions in color-mass space leads to a straightforward 
way to measure the color evolution of individual galaxies. We define two boundaries in color, $C_{\rm blue}$ and $C_{\rm red}$, 
such that a galaxy with $C < C_{\rm blue}$ is classified as blue, $C > C_{\rm red}$ as red, and $C_{\rm blue} < C < C_{\rm red}$ 
as an intermediate (`green valley') system. Each boundary is taken as some multiple of the width of that population away 
from its center location, both varying with stellar mass as well as redshift. Specifically, 

\begin{equation}
  \begin{aligned}
    C_{\rm blue}(M_\star,z) &= \mu_{\rm blue}(M_\star,z) + f_{\rm blue} \cdot \sigma_{\rm blue}(M_\star,z) \,,\\
    C_{\rm red}(M_\star,z)  &= \mu_{\rm red}(M_\star,z) - f_{\rm red} \cdot \sigma_{\rm red}(M_\star,z) \,.
  \end{aligned}
\end{equation}

\noindent The parameters $f_{\rm blue}$ and $f_{\rm red}$ set the precise locations of the color boundaries. We take 
$f_{\rm blue} = f_{\rm red} = 1.0$ as a compromise between (i) the separation between the red and blue populations, 
and (ii) contamination of the intermediate group by non-transitional galaxies. The timescales we measure below can be 
modulated by these values, which is to say, the time it takes galaxies to evolve through a region of color 
space depends on the definition of that region. 

For every galaxy at $z\,=\,0$ we follow the evolution of its $M_\star$ and (g-r) color back to $z\,=\,1$. We locate all 
crossings of the evolving red and blue boundaries, if applicable. For each galaxy we then know the first time (highest 
redshift) it left the blue population $t_{\rm blue}$, the time it subsequently joined the red population $t_{\rm red,ini}$, 
and the timescale of its color transition $\Delta t_{\rm green}$ -- its green valley occupation time. We also know its stellar 
mass upon exiting the blue population $M_\star(t_{\rm blue})$ and when entering the red population $M_\star(t_{\rm red,ini})$. 
Any mass growth after joining the red sequence is given by $\Delta M_{\rm \star,red} = M_\star(z=0) - M_\star(t_{\rm red,ini})$. 
After becoming red, a galaxy may at some later point drop out of the red population by satisfying $C < C_{\rm red}$, in which 
case we record the number $N_{\rm rejuv}$ and time(s) $t_{\rm rejuv}^i$ ($i = 0, 1, 2, ...$) of these `rejuvenation' events, 
as well as the mass growth $\Delta M_\star(t_{\rm rejuv}^i)$ and timescale(s) $\tau_{\rm rejuv}^i$ of each, if the galaxy 
rejoins the red population thereafter.

\begin{figure}
\centering
\includegraphics[angle=0,width=3.4in]
  {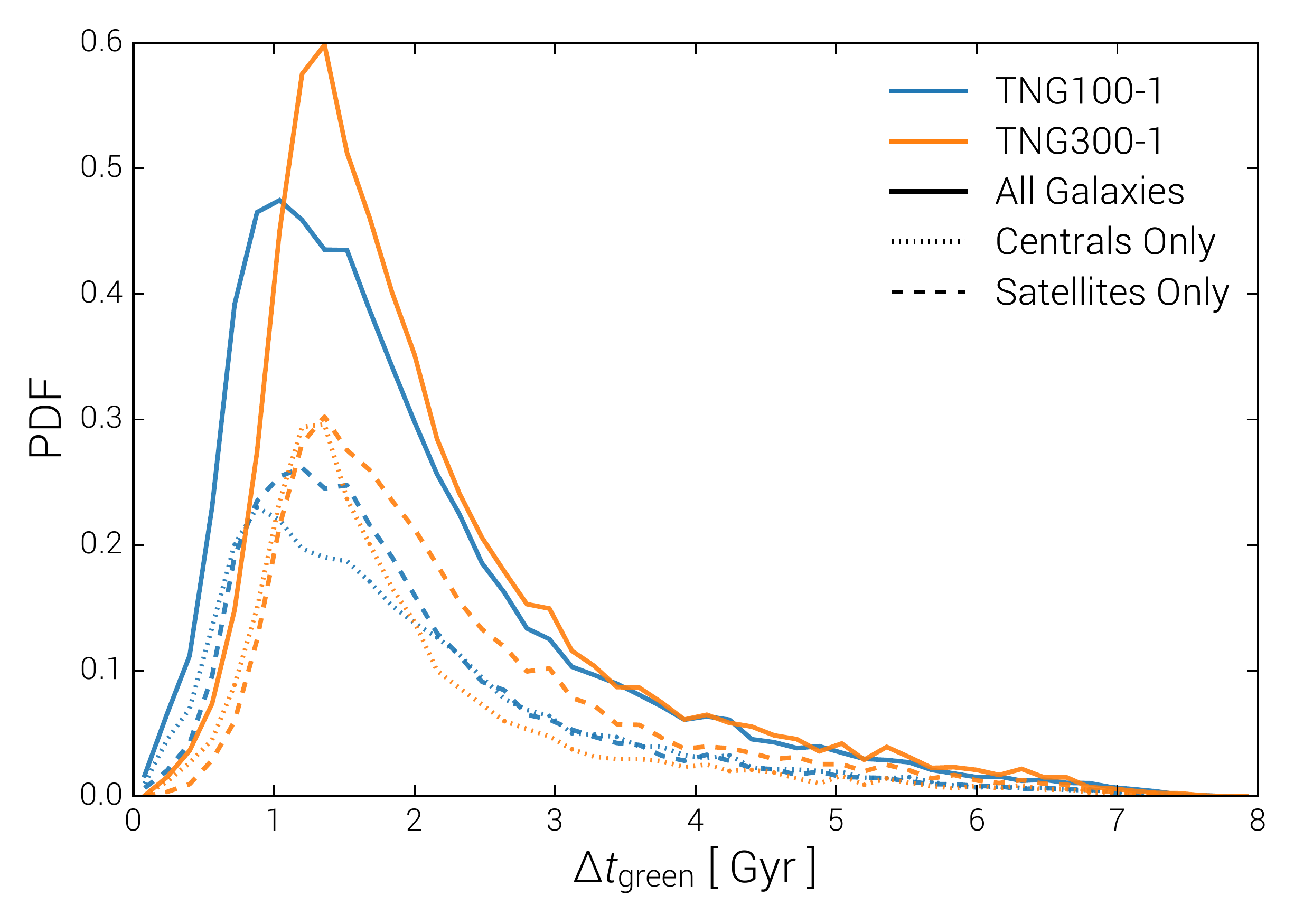}
\includegraphics[angle=0,width=3.4in]
  {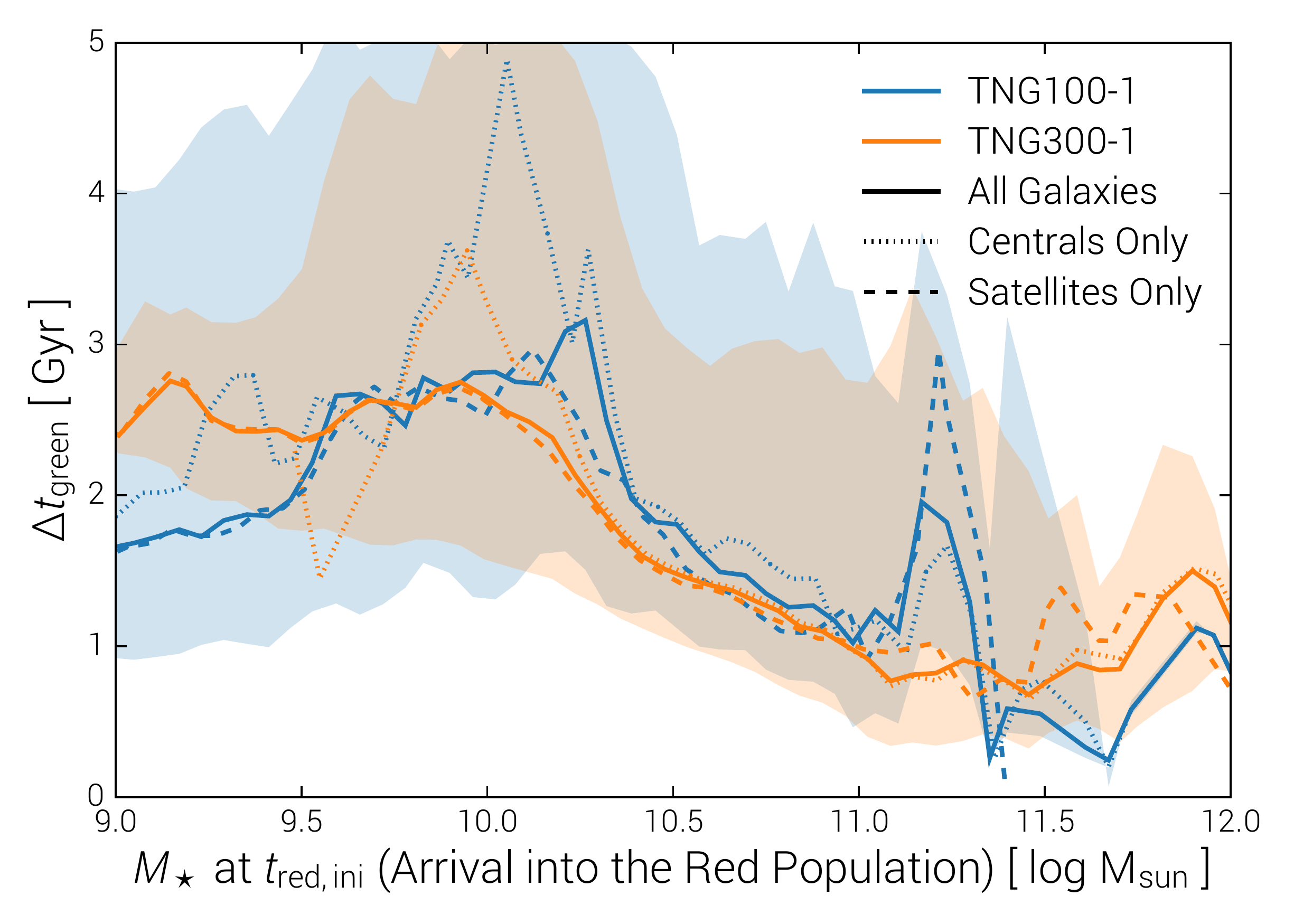}
\caption{ 
\textbf{Top.} The timescale of color transition $\Delta t_{\rm green}$ measured as $(t_{\rm red,ini} - t_{\rm blue})$, 
corresponding to the time difference between leaving the blue population and entering the red population. We show  
for TNG100/TNG300 the timescale PDF for all galaxies with $M_\star > 10^{9}$\,\msun (solid) and the separate distributions for 
$z\,=\,0$ central versus satellite galaxies only (dotted and dashed, respectively). In all cases the timescale 
distribution is broad, approximately log-normal, with a peak at $\sim$\,1.5\,Gyr.
\textbf{Bottom.} Relation between $\Delta t_{\rm green}$ and stellar mass when the galaxy 
became red $M_\star( t_{\rm red,ini} )$. At $M_\star \la 10^{10}$\,\msun\! galaxies require $\sim$\,2.5\,Gyr to transit 
the green valley. Above this mass, we find a decreasing trend with mass: galaxies which arrive into the red 
population with more stellar mass do so more rapidly. Shaded areas show the 10-90th percentiles for the 
`All Galaxies' samples.
 \label{fig_dt_green}} 
\end{figure}

\begin{figure*}
\centering
  \includegraphics[angle=0,width=3.4in]
    {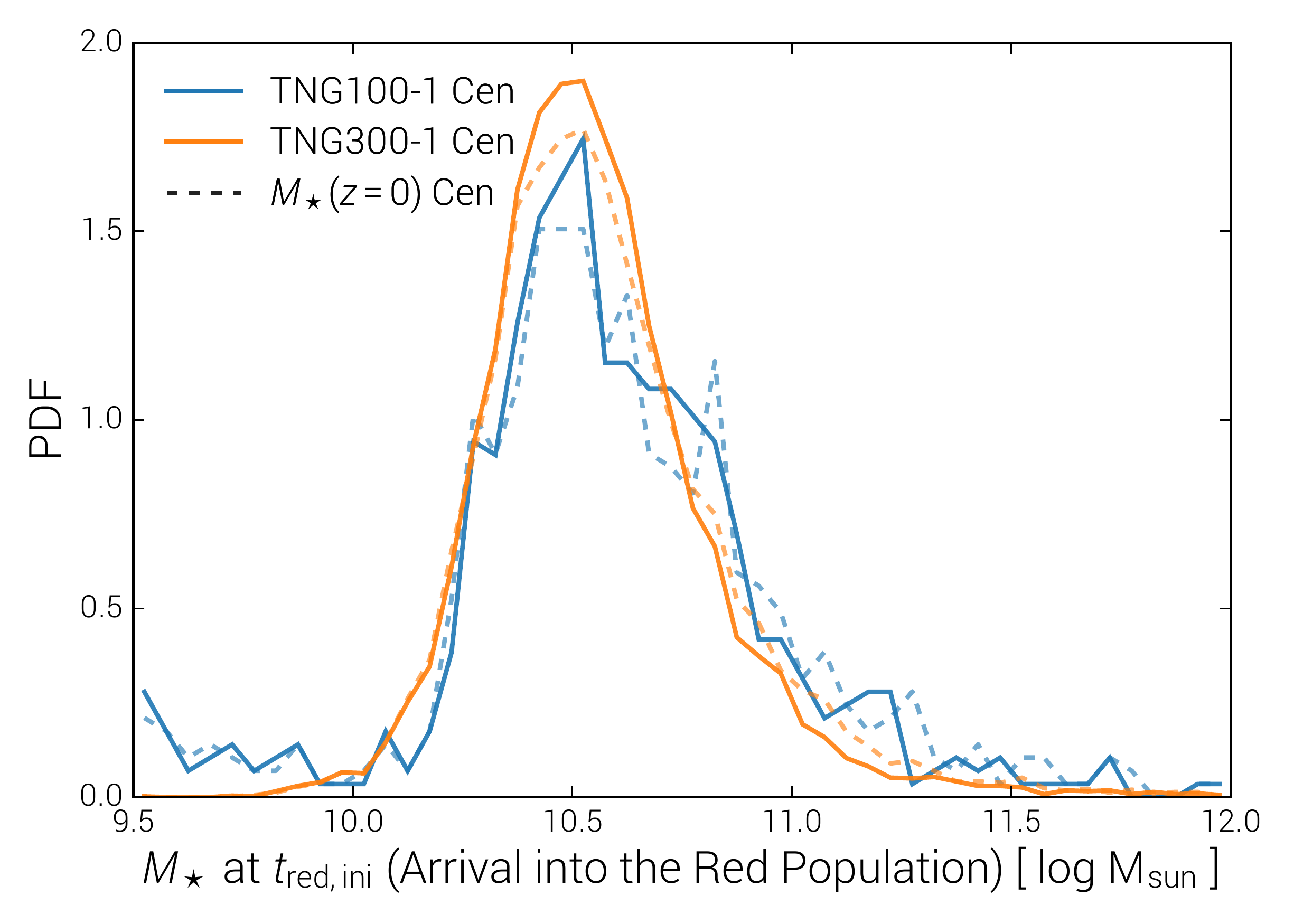}
  \includegraphics[angle=0,width=3.4in]
    {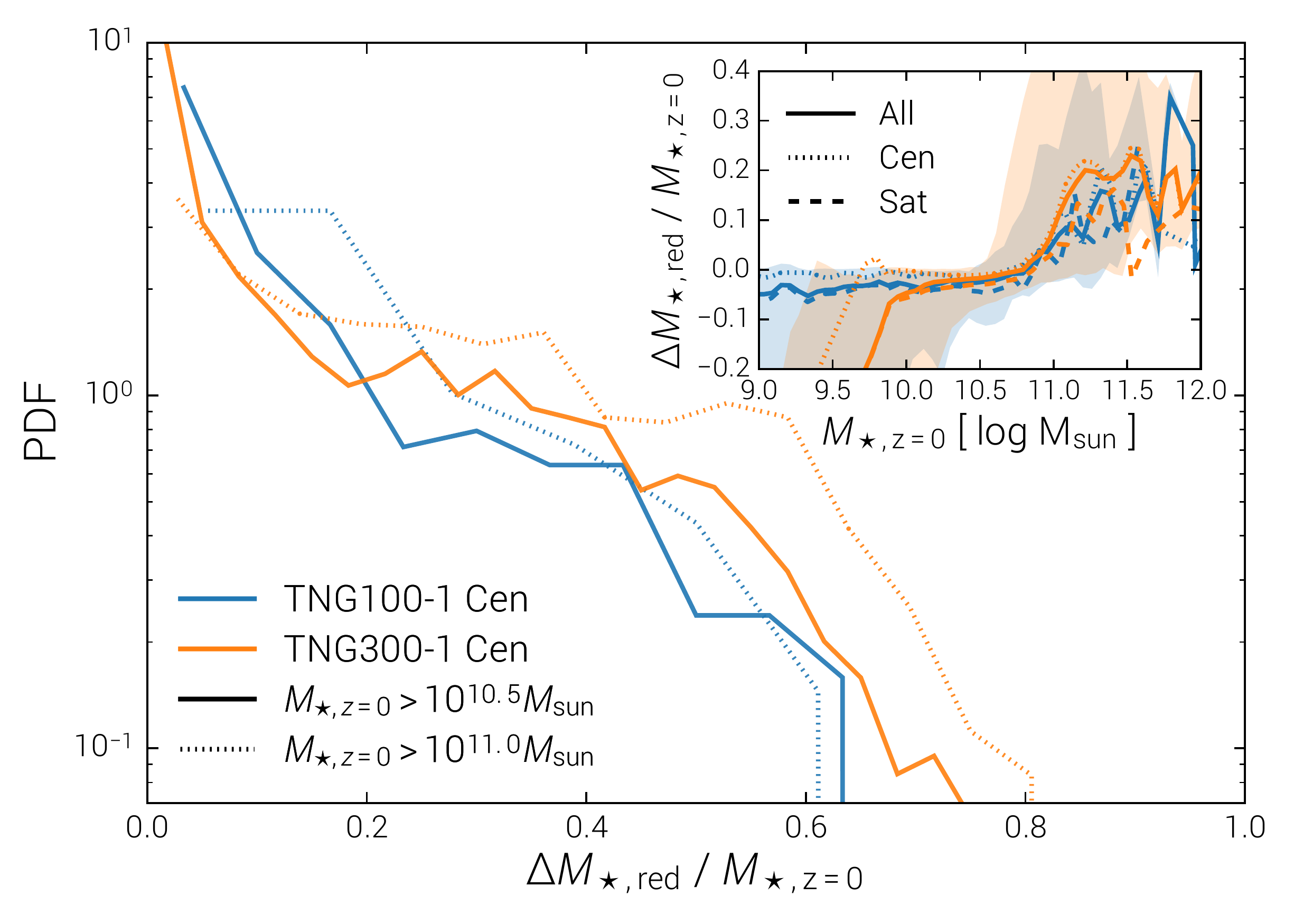}
\caption{
\textbf{Left.} The initial stellar mass of galaxies joining the red population, $M_\star(t_{\rm red,ini})$. Only 
\textit{central} galaxies at $z\,=\,0$ are shown to exclude the contribution from low-mass satellite reddening. In 
dotted lines we show the distributions of $z\,=\,0$ stellar masses of these same samples of galaxies, which are qualitatively 
quite similar.
\textbf{Right.} The fraction of final $z\,=\,0$ stellar mass of galaxies which originates from growth after joining 
the red population $\Delta M_{\rm \star,red} / M_{\rm \star,z=0}$. Only central galaxies with 
$M_{\rm \star,z=0}$ above the indicated minima are included to focus on the massive population. 
\textbf{Right Inset:} Correlation between this red-phase fractional mass growth and the $z\,=\,0$ stellar mass of the galaxy. 
Shaded areas show 10-90th percentiles for the `All' samples.
 \label{fig_Mstar_redini}}
\end{figure*}

Figure \ref{fig_dt_green} (top panel) presents the distribution of color transition timescales $\Delta t_{\rm green}$ for all 
galaxies with $M_\star > 10^{9}$\,\msun at $z=0$. Approximately 60\% of all galaxies transition between the blue and red populations 
in less than two billion years, with the distributions peaked at $\sim$\,1.5\,Gyr. An extended tail reaches to much longer 
timescales, with $\sim$\,10\% of galaxies taking 4\,Gyr or longer to transition in color. Although broad, the distributions 
of $\Delta t_{\rm green}$ are unimodal (roughly log-normal) and do not necessarily require the existence of multiple color 
transition pathways with distinct physical timescales. Quantitatively, in TNG100-1 the median $\Delta t_{\rm green}$ across 
all galaxies is 1.6\,Gyr with the 10-90th percentiles spanning 0.7\,Gyr to 3.8\,Gyr. 
There is only a minor difference between $z\,=\,0$ central versus satellite galaxies: satellites have a slightly narrower 
distribution of color transition timescales, P$_{10}^{90}$\,=\,[0.8\,Gyr, 3.7\,Gyr]. This reflects a non-trivial similarity 
between central and satellite quenching timescales, the latter arising only after some threshold of environmental influence.

In Figure \ref{fig_dt_green} (bottom panel) we 
show the timescale of color transition $\Delta t_{\rm green}$ as a function of the `initial red' galaxy stellar mass 
$M_\star(t_{\rm red,ini})$. Low mass systems which enter the red population with $M_\star \la 10^{10}$\,\msun\! take 
on average 2\,$-$\,3 billion years to become red (albeit with significant variation). Galaxies arriving with progressively 
larger stellar masses do so quicker, taking only $\sim$\,1\,Gyr at 
$M_\star \ga 10^{11}$\,\msun\!. We suggest that the increasingly common onset of low-state BH feedback at 
$\sim 10^{10.5}$\,\msun\! provides for the steep drop in $\Delta t_{\rm green}$ for centrals evident around this mass. 

Upon completion of a transition from blue to red over this timescale, galaxies will arrive to the red population with a 
range of stellar masses. Figure \ref{fig_Mstar_redini} shows the distribution of $M_\star(t_{\rm red,ini})$, the initial 
$M_\star$ upon first crossing $C_{\rm red}$ and entering the red population, for central galaxies (left panel). 
`Initial red' galaxy masses are peaked strongly at $\simeq$\,$10^{10.5}$\,\msun\!, and it would be 
tempting to conclude that the vast majority of massive galaxies join the red population at its low-mass end. However, 
the drop-off towards large $M_\star$ also encapsulates the correspondingly steep stellar mass function. Comparison with 
the $z\,=\,0$ stellar mass of these same populations (dashed lines) clarifies that most galaxies join the red population 
roughly in proportion to where they reside at $z\,=\,0$, and that we should examine the differential stellar mass growth.

Even in the complete absence of any further in-situ star formation, a newly red galaxy may still increase its $M_\star$ 
down to $z\,=\,0$. In Figure \ref{fig_Mstar_redini} (right panel) we show the distribution of this mass increase $\Delta M_{\rm \star,red}$ 
normalized by final $z\,=\,0$ stellar mass (main panel). We include only central galaxies with 
\mbox{$M_{\rm \star,z=0}$\,$>$\,$10^{10.5}$\,\msun\!} to focus on the massive population and exclude the negative tail 
arising from satellite mass loss. In this regime \mbox{0\,$<$\,$\Delta M_{\rm \star,red} / M_{\rm \star,z=0}$\,$<$\,1} and 
for TNG100-1 the average fraction of the final stellar mass which accumulates post-quiescence is $0.11$, while the upper 
10th percentile of galaxies exceed $0.41$. Above 10$^{11}$\,\msun\! these values increase to $0.25$ and $0.53$, respectively. 
Here, $\simeq$\,18\% of all galaxies acquire half or more of their present day mass on the red sequence. A small number 
of the most massive galaxies can acquire nearly all of their $z=0$ stellar mass while on the red sequence - we find examples 
as high as 80\%. The inset panel (upper right) quantifies this correlation between red-phase mass growth and total stellar mass. 
The observed differences between TNG100 and TNG300 primarily reflect the lower mass resolution of the latter.

We note that for satellite galaxies with $M_\star$ below 10$^{10.5}$\,\msun\! the mean $\Delta M_{\rm \star,red} / M_{\rm \star,z=0}$ 
is $-0.22$ with the lowest 10th percentile obtaining $-0.44$. The lowest mass satellite galaxies can lose $>$\,50\% of their 
stellar mass \textit{after} becoming red. Post-reddening stellar mass growth (or loss) is clearly a strong function of both stellar 
mass and environment.


\section{Discussion} \label{sec_discussion}

\begin{figure*}
\centerline{\includegraphics[angle=0,width=6.8in]{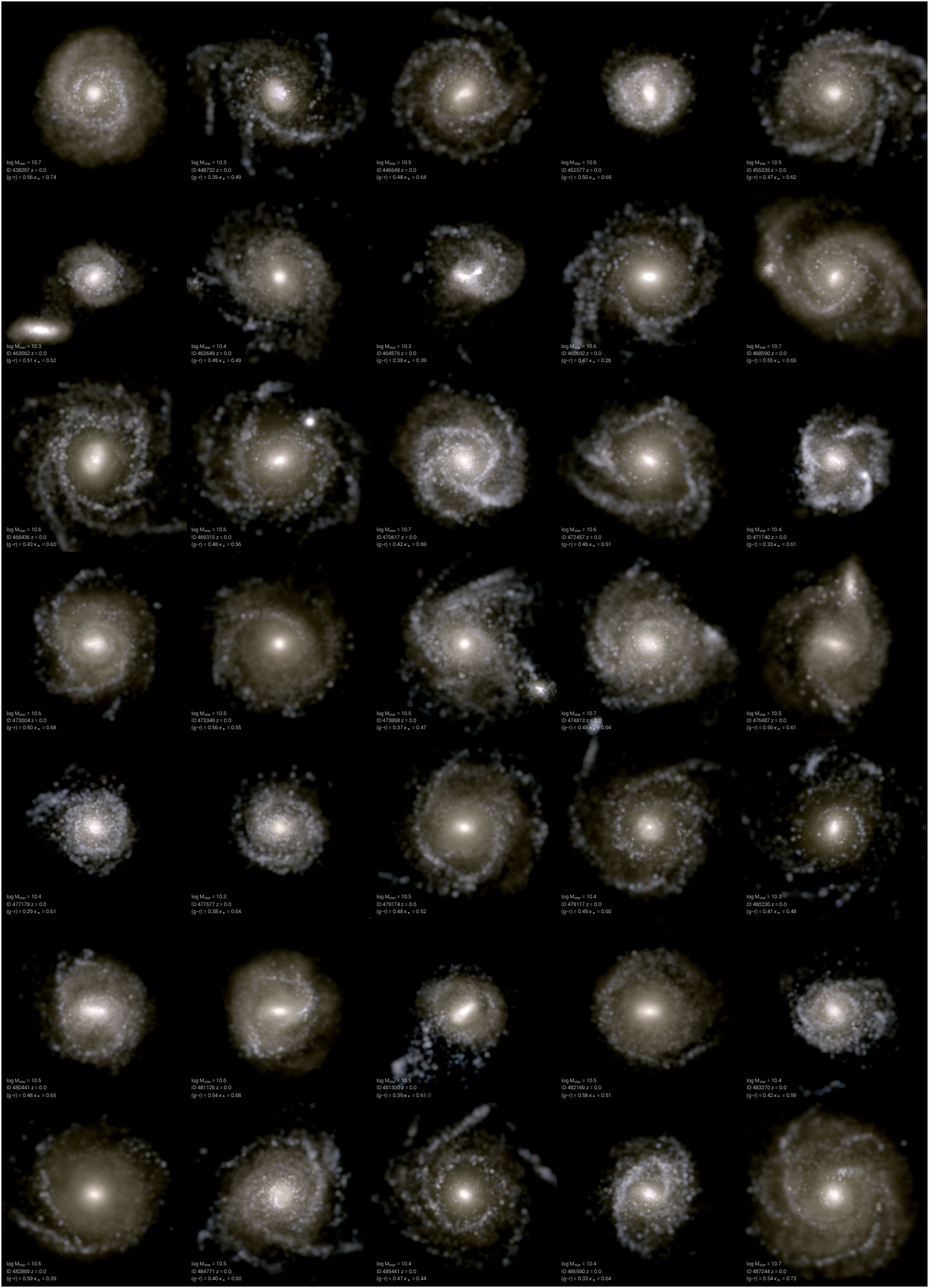}}
\caption{ Stellar light composites of a sample of $z\,=\,0$ \textit{blue} galaxies, selected as having (g-r)\,$<$\,0.6, 
and taken from the halo mass bin \mbox{$10^{12}$\,$<$\,$M_{\rm halo}$/M$_\odot$\,$<$\,$10^{12.2}$}. Imaged in NIRCam 
f200W, f115W, and F070W filters (face-on), no dust. Every panel is 60\,kpc on a side.
 \label{fig_stamps_blue}} 
\end{figure*}

\begin{figure*}
\centerline{\includegraphics[angle=0,width=6.8in]{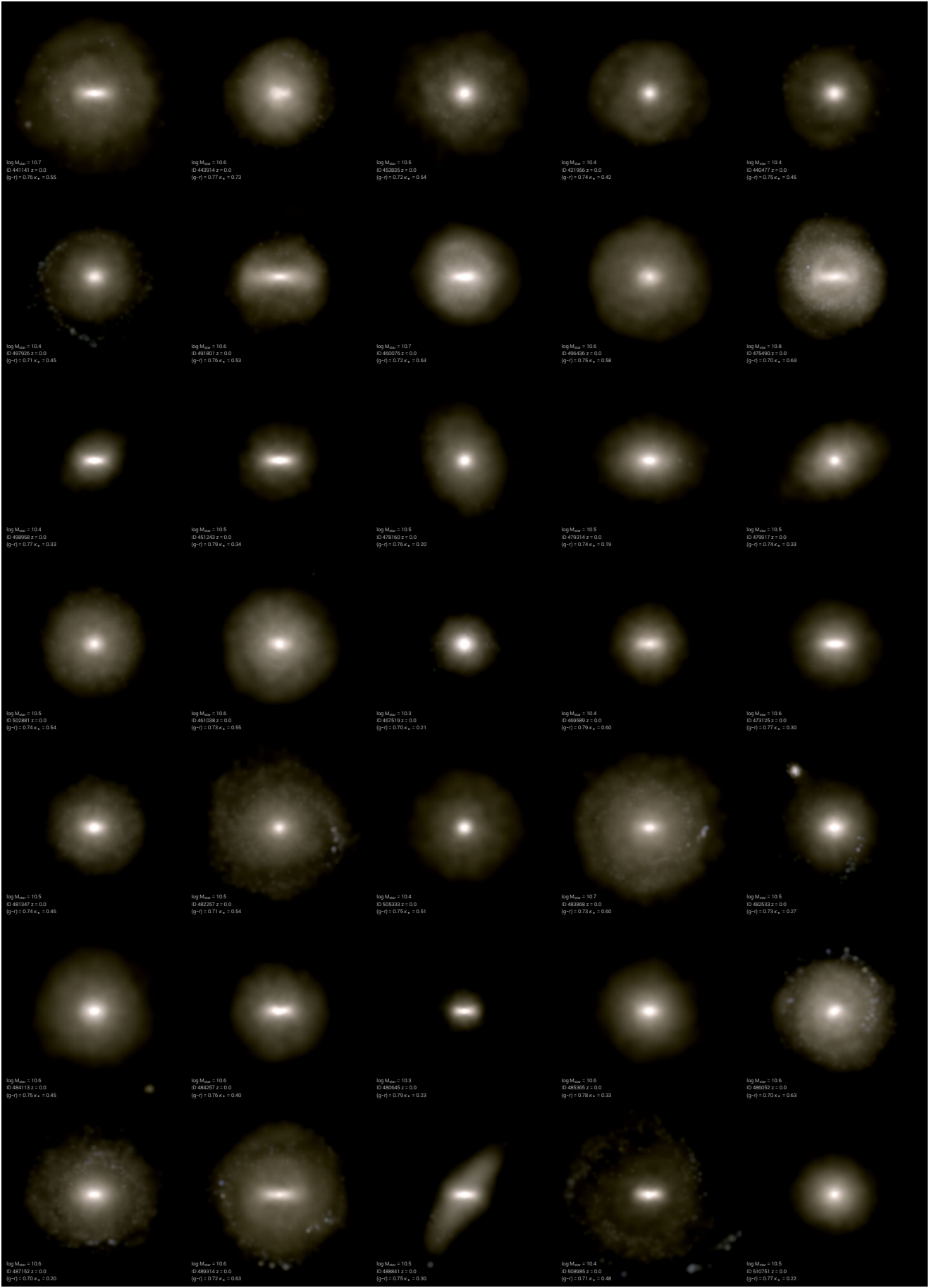}}
\caption{ Stellar light composites at $z\,=\,0$ as in Figure \ref{fig_stamps_blue} except here for \textit{red} galaxies, 
selected as having (g-r)\,$>$\,0.6 and in the same halo mass bin \mbox{$10^{12}$\,$<$\,$M_{\rm halo}$/M$_\odot$\,$<$\,$10^{12.2}$}, 
corresponding to the low-mass end of the red sequence.
 \label{fig_stamps_red}} 
\end{figure*}

So far we have presented our analysis of the color distribution of TNG galaxies without ever showing either a blue or 
red galaxy. We therefore turn to Figures \ref{fig_stamps_blue} and \ref{fig_stamps_red} which visualize two $z\,=\,0$ samples 
of 35 blue and 35 red galaxies, respectively, making the connection to stellar morphology.

Stellar light composites in the optical emphasize younger stars in the blue channel and older stars in the red. Galaxies 
are shown face-on, projections are 60 physical kiloparsecs on a side\, and each panel includes the galaxy stellar mass, 
subhalo ID, \mbox{(g-r)} color, and $\kappa_{\rm \star,rot}|_{J_z>0}$, a proxy for the rotational support  
\citep[strong spheroids having $\kappa \la 0.3$ and strong disks having $\kappa \ga 0.6$;][]{rodriguezgomez17}. Both 
samples are drawn from a halo mass bin \mbox{$10^{12}$\,$<$\,$M_{\rm halo}$/M$_\odot$\,$<$\,$10^{12.2}$} where overlapping 
red and blue populations exist at similar stellar masses. Here we select blue galaxies as having \mbox{(g-r)\,$<$\,$0.6$}, 
which may also include transitional color objects, while we require \mbox{(g-r)\,$>$\,$0.6$} for the red galaxy sample.

In this roughly Milky Way mass bin the bluer galaxies are typically large, star-forming disks. Spiral features are often 
present, commonly traced by the blue light of younger stellar populations. Spiral arms range from tightly wound 
configurations, to open grand-design patterns, to flocculent and even more irregular arrangements, although we caution that 
the we do not expect the physical mechanisms of spiral arm formation to be realistically captured in cosmological simulations 
like TNG. Some systems have strong central bulges, bar-like features, clumpy disk sub-structures, and/or merging companions; 
$10.4$\,$\la$\,$M_\star$\,/\,\msun\!\,$\la$\,$10.6$ and $0.5$\,$\la$\,$\kappa_\star$\,$\la$\,$0.7$ are typical values.

For passive centrals this is a relatively low mass scale, and the red galaxy sample is a mix of spheroidal and disk 
morphologies. We see that some `red disks' exist, as well as the more common featureless ellipticals. Many of these systems 
exhibit fine-grained phase space structures including stellar shells and elongated stellar streams from tidally destroyed 
satellites, indicative of recent merger activity -- both features become more common at higher $M_\star$. Typically 
$10.5$\,$\la$\,$M_\star$\,/\,\msun\!\,$\la$\,$10.7$ and $0.2$\,$\la$\,$\kappa_\star$\,$\la$\,$0.4$ for the spheroids, 
while the passive disks continue to have $\kappa_\star$\,$\ga$\,$0.6$.

\restoregeometry
\addtolength{\topmargin}{-0.8cm}
\addtolength{\textheight}{1.6cm}

\restoregeometry

These two visual samples demonstrate the existence 
of multiple, distinct galaxy types spanning a wide range of stellar morphologies and structural characteristics. This 
diversity of the galaxy population \textit{even at fixed mass}, as observed to occur in the real universe, represents a 
fundamental achievement (and stringent requirement) of any theoretical model for galaxy formation.

\begin{figure*}
\centerline{
\includegraphics[angle=0,width=6.4in]{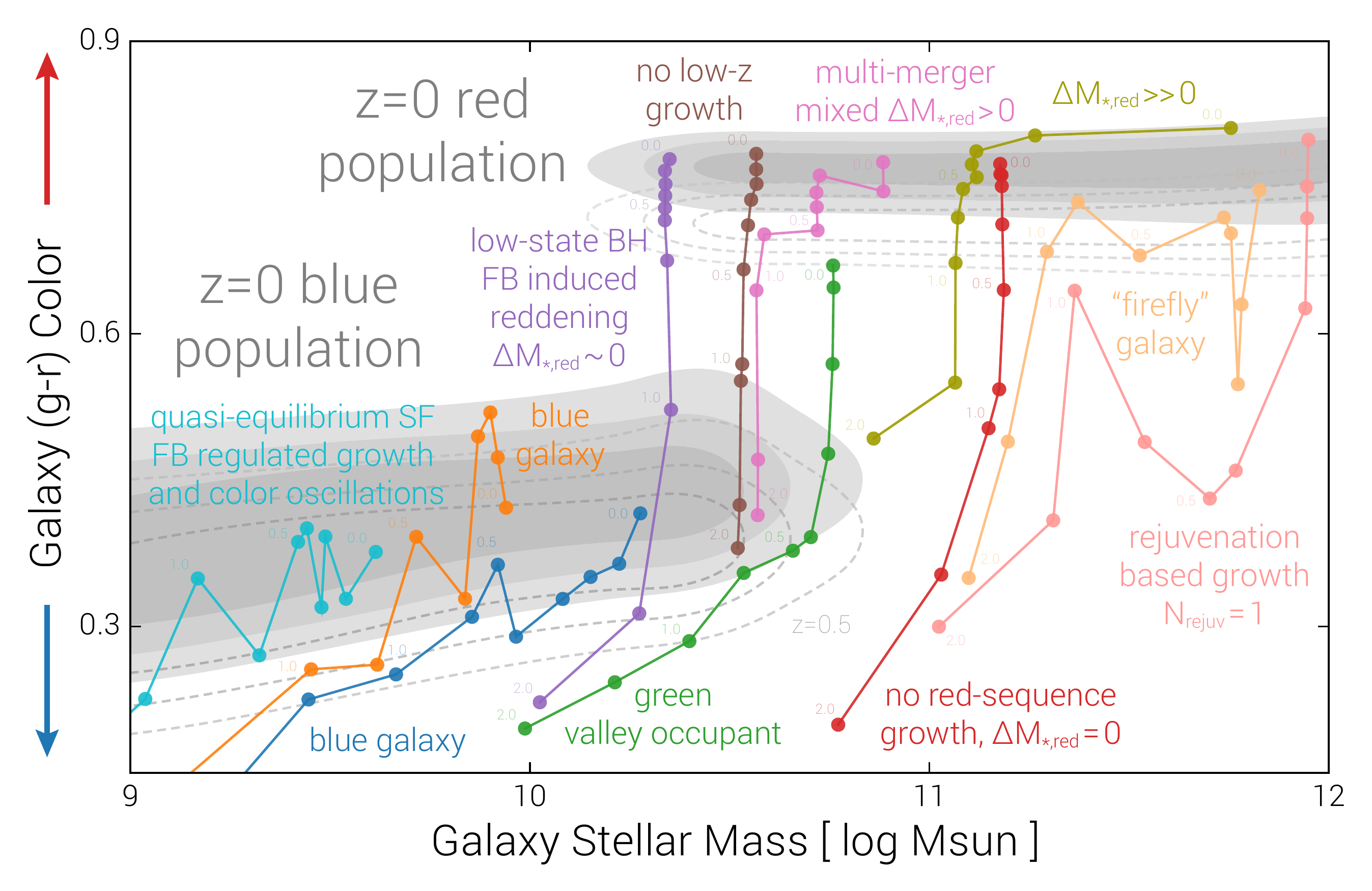}}
\caption{ Schematic diagram of galaxy evolution across the color-mass plane. The distribution of all $z\,=\,0$ galaxies 
is shown by the gray contours, while the dotted lines indicate the bulk evolution of the red and blue populations to 
$z\,\simeq\,0.5$, both moving downward in (g-r). Eleven colored lines show the \textbf{characteristic evolutionary pathways 
of individual, central galaxies}, from at most $z\,=\,2$ to the present day. Although drawn from actual tracks in TNG100-1, 
we intend them as prototypical demonstrations of the various trajectories of central galaxies towards $z\,=\,0$ and in 
particular the ways they can ascend to the red population (see text).
 \label{fig_schematic}} 
\end{figure*}

\subsection{Galaxy Color Evolution and the Formation of the Color-Mass Plane}

We seek to draw together our conclusions and thereby arrive at a coherent picture for the color distribution of low redshift 
galaxies and the physical mechanisms most responsible. In Figure \ref{fig_schematic} we therefore present a schematic view 
of the color-mass plane and the movement of galaxies therein. Gray contours show the distribution of color versus stellar 
mass for all galaxies at redshift zero, which divides broadly into the blue population at low stellar mass and the red 
population at high stellar mass. The dotted contours indicate the bulk evolution of these two groups to $z\,\sim\,0.5$, each 
becoming bluer at earlier times. With eleven colored lines we show characteristic evolutionary tracks which individual 
\textit{central} galaxies can take through this space. They encompass all prototypical behaviors without being exhaustive.

The red population continually incorporates new members, as galaxies transitioning over the timescale $\Delta t_{\rm green}$ 
arrive with a stellar mass $M_\star( t_{\rm red,ini} )$. Figure \ref{fig_Mstar_redini} quantified the distribution of this 
`initial red' mass, which is strongly peaked at $\sim$\,$10^{10.5}$\,\msun\! -- the purple and pink tracks which ascend 
almost vertically from the high-mass end of the blue population are therefore typical in this regard. Although most redshift 
zero red sequence galaxies, by number, join the population at its low-mass end, this is by and large a reflection of the steep 
stellar mass function -- most $z\,=\,0$ red sequence galaxies also simply reside at its low-mass end.

Even without any further in-situ star formation, passive red galaxies may grow further in $M_\star$ down to $z\,=\,0$ by 
acquiring new stars through mergers \citep{vandokkum99,naab06}, with accreted stellar mass fractions of 50\% or higher for 
galaxies exceeding 10$^{11}$\msun \citep{rodriguezgomez15,pillepich17a}. The multi-merger track (pink) and $\Delta M_{\rm \star,red} \gg 0$ 
track (gold) respectively provide typical and extreme examples of this mechanism. The former undergoes two distinct dry merger events 
which together increase the galaxy stellar mass by a factor of two, with marginal impact on its color. The latter undergoes 
several dry mergers which increase its mass by a factor of four, becoming more red in the process. Figure \ref{fig_Mstar_redini} 
quantifies this red sequence mass growth in terms of $\Delta M_{\rm \star,red} / M_{\rm \star,z=0}$, the fraction of the 
redshift zero stellar mass of a galaxy accumulated after its entrance into the red population -- above 
$M_\star$\,$>$\,10$^{11}$\,\msun\! galaxies on average accumulate 25\% of their final mass after reddening.

We naturally contrast our findings for $\Delta M_{\rm \star,red}$ with the picture proposed in \cite{faber07}. In particular, 
their Figure 10 presents three schematic possibilities of galaxy migration to, and growth along, the red sequence. Adopting 
that nomenclature, we here specifically refer to galaxies which enter the red population at its low mass end ($\sim$\,10$^{10.5}$\,\msun\!) 
as `early quenching' and those which enter the red population at higher masses as `late quenching'. We rule out the 
`early quenching + dry merging' scenario. Namely, while early quenchers are plentiful (epitomized in our purple and brown 
tracks), they grow the least through subsequent dry stellar accretion. The `late quenching + no dry merging' scenario is 
common and important. We call these $\Delta M_{\rm \star,red} = 0$ tracks, typified by our red trajectory. We note that 
evolutionary paths of this shape (with $\Delta M_{\rm \star,red} \simeq 0$) occur across the full spectrum of 
$M_\star( t_{\rm red,ini} )$, and at any redshift. The `mixed, quenching + dry merging' scenario is also common and important. 
We label these as $\Delta M_{\rm \star,red} > 0$ tracks, typified by the pink and gold trajectories. Both mechanisms are 
present; therefore, neither alone can capture the manner in which the red sequence builds up from the entire galaxy population.

Our contribution to this theoretical picture is given in Figure 
\ref{fig_Mstar_redini} (inset), which quantifies that high mass galaxies experience more red-phase growth. That is, 
in the `mixed' scenario of \cite{faber07}, arrows denoting $\Delta M_\star$ along the red sequence become longer towards 
higher redshift zero $M_\star$. In general, we also find that galaxies which enter the red population with more mass do so 
later: at $10^{10.5}$\,\msun\! systems transition at a mean $\bar{z}_{\rm red,ini}$\,$\sim$\,0.6 while those which enter at 
$10^{11.5}$\,\msun\! do so at $\bar{z}_{\rm red,ini}$\,$\sim$\,0.3.

As we have already seen, there is significant variation in $\Delta t_{\rm green}$ even at similar stellar mass. Figure 
\ref{fig_dt_green} quantifies the distribution of this green valley occupation timescale, and we find a median value 
across all galaxies of 1.6\,Gyr with the 10-90th percentiles spanning 0.7\,Gyr to 3.8\,Gyr. A large fraction of the 
population is therefore consistent with passive reddening after a short quenching timescale, consistent with observational 
demographics of post-starburst galaxies \citep[e.g.][]{wild16} and passive evolution times using single SSPs \citep[e.g.][]{trayford16}. 
At the same time, the presence of existing older stellar population could induce a more rapid blue to red transition, while 
dust may also be an important factor in color transition timescales \citep[e.g.][]{gutcke17}.
As the symbols for each track 
in our schematic figure are placed at $z = \{0, 0.1, 0.2, 0.3, 0.4, 0.5, 0.7, 1.0, 1.5, 2.0\}$, they are spaced roughly 1 
Gyr apart. As an example, while the brown and pink paths begin at the same mass and color at $z\,=\,2$, they traverse the 
green valley over different timescales, the multi-merger case requiring $\sim$\,2\,Gyr less. \cite{trayford16} also computed 
color transition timescales from cosmological simulations, finding a median of $\Delta t_{\rm green} \simeq 2.0 - 2.5$ Gyr 
for $10^{10}$\,$<$\,$M_\star$\,/\,\msun\,$<$\,$10^{11}$ with a tail towards longer values and no strong mass dependence. 
Despite being qualitatively similar, in detail this result differs from our findings in Figure \ref{fig_dt_green} 
by up to a factor of two, as well as in the decreasing dependence of $\Delta t_{\rm green}$ 
on stellar mass. This is undoubtedly due to a combination of different physical models and so quenching mechanisms between 
the two simulations, compounded with different definitions of the color transition region and the definition of the transition 
between blue and red galaxies. We find that the most massive 
central galaxies of the present epoch have spent the least amount of time in the green valley, 
partially reflecting the merging loci of the blue and red populations towards the highest masses.

Although we have not presented comprehensive results on the process of rejuvenated star formation in the red population, 
we make a few brief observations from TNG100-1. Once red, the majority of galaxies do not ever leave the red population. 
Across the entire resolved mass range, 94\% of redshift zero galaxies have $N_{\rm rejuv}$\,=\,0. If we restrict our 
attention to massive, passive galaxies by considering $M_{\rm \star,z=0}$\,/\,M$_\odot$\,$>$\,$10^{11}$\msun\! we find 
that 90\% of galaxies have never undergone a rejuvenation event, while 10\% have rejuvenated once and $\sim$\,1\% have 
gone through two or more such events. These `firefly' galaxies (light orange line) continually relight their star 
formation and so jump along the red sequence by increasing their stellar mass in intermittent flashes. 
We find a maximum of $N_{\rm rejuv}$\,=\,5, although such extreme cases represent a negligible fraction of the galaxy 
population. The distribution of rejuvenation timescales $t_{\rm rejuv}^i$ has a similar shape as the $\Delta t_{\rm green}$ 
distribution, with a slightly lower normalization. The median time between exiting and re-entering the red population is 
1.1\,Gyr with a percentile spread P$_{10}^{90}$\,=\,[0.5\,Gyr, 2.7\,Gyr]. The duty cycle of rejuvenation in massive red 
galaxies should correspond to observable residual star formation in such objects \citep{fogarty15,mcdonald16}. Some massive 
galaxies can attribute almost all of their late-time stellar growth to a single large rejuvenation event. The light red track 
provides one example -- despite the fact that this galaxy enters the red population already by $z\,=\,1$ it resumes star 
formation and does not finally quench until $\sim$\,5\,Gyr later. Note that the total $\sum \Delta M_\star(t_{\rm rejuv}^i)$ 
is included by definition in the $\Delta M_{\rm \star,red}$ values we present throughout, and that we could miss rejuvenation 
events due to the finite time spacing of our simulation snapshots or due to the precise definition of the red color boundary.

\subsection{The Connection to Galaxy Morphology}

We have also not addressed the question of morphological transformation, an intriguing aspect of galaxy formation 
which is intimately connected to color. Specifically, we have not attempted to discern if color transition and the 
quenching of star formation is always accompanied by morphological change, and if so, if these occur co-temporal or if one 
precedes the other. From our blue/red samples in Figures \ref{fig_stamps_blue} and \ref{fig_stamps_red} we already see by eye 
the relation between morphology and color in this halo mass bin \citep[see also][]{snyder15}. We add a few brief comments 
here. At fixed $M_\star = 10^{10.5}$\,\msun\! we find a trend of morphology as measured through $\kappa_\star$ and 
\mbox{(g-r)} color, such that redder galaxies are on average less rotationally supported. We find no correlation between 
$\kappa_\star(z=0)$ and $\Delta t_{\rm green}$. On the other hand, $\kappa_\star$ exhibits a suggestive scaling with 
red-phase mass growth, such that $\kappa_\star \simeq 0.4$ for $\Delta M_{\rm \star,red} / M_{\rm \star,z=0} \sim 0$ while 
$\kappa_\star \la 0.2$ for $\Delta M_{\rm \star,red} / M_{\rm \star,z=0} \ga 0.25$. That is, galaxies which acquire more 
mass on the red sequence are driven towards preferentially spheroidal morphologies; the underlying physical cause is 
undoubtedly merger activity, which lowers the former while increasing the latter. A similar trend exists with $\Delta M_{\rm green}$, 
hinting that mass growth in the green valley is associated with disturbed, non-steady-state star formation.
This may indirectly agree with the observationally based hypothesis of \cite{powell17} that a lack of green spheroids results 
from the rapidity of their quenching via mergers. However, the lack of correlation between $z\,=\,0$ morphology and 
$\Delta t_{\rm green}$ does not appear to support the idea that early-type (or late-type) galaxies should arrive to the red 
population over characteristically shorter (or longer) transition timescales due to either near instantaneous quenching or 
slow secular gas exhaustion, respectively \citep{schawinski14}. A more detailed study along these lines is clearly needed.
The interwoven pathways of stellar reddening and structural change, together with the origin of red disks, blue spheroids, and 
the connection with central stellar bulges, all promise to be a fruitful future direction of investigation. 

\subsection{Final Considerations}

Throughout this paper we have intentionally focused on the \textit{massive} red population and its emergence from previously 
star-forming galaxies. That is, we have neglected the properties and evolution of red satellites, satellite quenching, and 
environment. The upper left corner of Figure \ref{fig_schematic} is populated by low-mass red galaxies, which are almost 
entirely satellites in dense environments. From observations we expect strong environmental dependence (including one-halo 
conformity) in color related properties such as color-split luminosity functions or stellar mass functions \citep{baldry06}, 
red fraction \citep{peng10}, color-split clustering \citep{wang12,farrow15,zu17,lawsmith17}, and quiescent fraction 
\citep{cucciati17}. We also expect different physical processes such as tidal and ram-pressure stripping to play important 
roles in satellite galaxy transformation, not only in color but also in star formation rate, gas content, and morphology 
\citep[e.g.][]{ann08,sales15}. We leave for the future an investigation of the environmental dependency of galaxy color in TNG.

Finally, we have not specifically addressed the balance between stellar feedback -- that is, our supernova driven galactic 
winds -- and blackhole feedback \citep[for an interesting view see][]{bower17}. We have shown that the low-state BH feedback mode is 
responsible for inducing color transformation in the TNG model and setting its characteristics. Interestingly, we find only a 
moderate correlation of blackhole mass with galaxy color at our canonical stellar transition mass of 
$10^{10.5}$\,\msun\!. Here the reddest galaxies have blackholes which are $\simeq$\,0.4\,dex more massive than 
their bluest counterparts. This is because blackhole mass growth in the TNG model is dominated by accretion in the high 
state (quasar mode). Therefore, even though transition to the low-state quenches galaxies and transforms galaxies from blue 
to red, it cannot leave a strong impression in the relic BH total mass. We do not see the same magnitude signature of excess SMBH 
mass as found in \cite{trayford16} (their Figure 3). However, in the integrated ratio of mass growth between the low and 
high states, the integrated ratio of energy injection (as shown in our Figure \ref{fig_bh_cumegy}), or the total energy injection 
in the low-state, we do see a clear marker of blackhole induced quenching. For example, at $M_\star = 10^{10.5}$\,\msun\! blue 
galaxies have a total integrated $E_{\rm injected,low} \simeq 10^{58}$\,erg, while red galaxies have injected a hundred 
times more energy in the low accretion-state, with median $E_{\rm injected,low} \simeq 10^{60}$\,erg. There is a pronounced 
increase in these BH related indicators at a color threshold of \mbox{(g-r) = 0.55 mag}.

In Figure \ref{fig_schematic} the remaining three lowest mass evolutionary tracks -- cyan, orange, and blue -- are all below 
this color threshold, and all still members of the blue population at $z\,=\,0$. Each has grown in stellar mass through 
quasi-equilibrium oscillations of star formation activity, which drives their optical color periodically up and down. Indeed 
the self-regulated nature of star formation should establish the color width of the blue population. Here we identify the 
origin of the features seen at low $M_\star$ in Figure \ref{fig_cmplane_fluxes}. Because of these oscillations, galaxies 
below a center line will have generally arrived from redder colors, while those above will have arrived from bluer colors. 
This establishes the separatrix in the streamline field and the butterfly pattern of the flux of galaxies through this region 
of the color-mass plane.


\section{Summary and Conclusions} \label{sec_summary}

In this paper we have introduced the TNG100 and TNG300 simulations, the first two efforts of the IllustrisTNG project, a series 
of `next generation' cosmological magnetohydrodynamical simulations using the updated, comprehensive TNG galaxy formation model. 
Leveraging their unique combination of volume, resolution, and physical fidelity, we model the optical colors of 
the simulated galaxy populations, incorporating a resolved treatment of attenuation from interstellar dust.\footnote{All the 
data generated herein, including the galaxy luminosities and colors with the varying dust models, will 
be made publicly available coincident with the future TNG public data release \citep[following][]{nelson15b}.} We first 
compare to observational data from SDSS and then, bolstered by the excellent agreement, study the relationships 
between color and other galaxy properties, the process of color transformation, and the physical mechanisms driving the 
buildup of the color-mass plane. Our main results are:

\begin{itemize}
\item The simulated (g-r) colors of TNG galaxies at low redshift are in excellent agreement with a quantitative comparison to observational 
      data from SDSS at $z\,<\,0.1$. The color bimodality for intermediate mass galaxies is demonstrably improved with respect 
      to the original Illustris simulation. We recover the locations in color of both the red and blue populations as a function of 
      $M_\star$, as well as the relative strength between the red and blue distributions, the location of the color minimum 
      between the two, and the location of the maximal bimodality in stellar mass. We show the same level of quantitative 
      agreement exists in other colors including \mbox{(u-r)}, \mbox{(u-i)}, while slightly less favorable comparison to the 
      dust-sensitive \mbox{(r-i)} may indicate shortcomings in our relatively simple dust modeling.
\item We use a double gaussian mixture model to characterize the red and blue populations and apply the same MCMC fitting 
      procedure to both the simulations and observations. The loci of the red and blue distributions agree across 
      \mbox{10$^9$\,$<$\,$M_\star$\,/\,\msun\,$<$\,10$^{12}$} to within 0.05 mag, while the scatter is consistent within 
      0.02 mag for blue galaxies and 0.01 mag for red galaxies. This foundational agreement allows us to identify 
      a second order discrepancy of possible interest: the slope of the red sequence may be too flat, an effect likely 
      driven by the finite aperture choice together with non-zero color gradients in extended stellar light.
\item As the two primary ingredients of colors beyond reddening, we directly compare the stellar ages and metallicities in TNG versus 
      observed relations. The luminosity-weighted, fiber aperture restricted stellar ages are in remarkable 
      agreement. The trend of $Z_\star$ versus $M_\star$ appears too shallow, implying that the 
      simulated stellar metallicities are too high for low-mass galaxies. However, we show that this discrepancy plausibly 
      arises only due to an overly simplistic comparison, and that a more robust analysis based on consistent fits to mock 
      SDSS fiber spectra is required (\textcolor{blue}{Nelson et al. in prep}).
\item Relating redshift zero (g-r) color to other galaxy properties, we find as expected that red galaxies are quiescent, 
      old, and gas depleted. Compared to otherwise increasing trends with $M_\star$, they also have lower gas-phase metallicities 
      and lower interstellar magnetic field strengths. We find a non-monotonic correlation, at fixed stellar mass, between 
      galaxy color and the $\beta$ ratio of magnetic to thermal pressure in the gaseous halo, such that the 
      $P_{\rm B}/P_{\rm gas}$ ratio peaks for green valley systems and drops sharply thereafter for red galaxies. We 
      point out an explicit prediction of the TNG model: at $M_\star \simeq 10^{10.5}$\,\msun\! blue galaxies should 
      have interstellar magnetic field strengths of $10-30$\,$\mu$G, while red galaxies at the same mass should have 
      much lower $|B_{\rm ISM}| \la 1$\,$\mu$G.
\item We attribute the primary driver of galaxy color transition in the TNG model to supermassive blackhole feedback in its 
      low-accretion state. We demonstrate that the ratio of the cumulative energy injection by blackholes in the high (quasar) 
      versus low (kinetic wind) state drops sharply around a characteristic $M_\star \sim 10^{10.5}$\,\msun\!, a value set by our 
      $M_{\rm BH}$-dependent accretion rate threshold. Having been fixed during model development in order to match the 
      stellar mass content of halos and in particular the knee of the stellar mass function, we conclude that color transition 
      at this mass scale is inseparably linked to a realistic high-mass galaxy population.
\item Across all $z\,=\,0$ galaxies, we measure a median color transition timescale $\Delta t_{\rm green}$ of $\sim$\,1.6\,Gyr, with a 
      10-90th percentile spread of $0.7 - 3.8$\,Gyr. This distribution of green valley occupation times is unimodal, 
      approximately log-normal, and does not show direct evidence for more than one characteristic timescale. We find 
      a relation between $\Delta t_{\rm green}$ and stellar mass, such that galaxies which enter the red population at larger 
      $M_\star$ do so with a correspondingly more rapid color transition and shorter green valley occupation.
\item We measure the amount of `red-phase' stellar mass growth, between the time a galaxy enters the red population and $z\,=\,0$.
      Massive galaxies with \mbox{$M_\star$\,$>$\,$10^{11}$\,\msun\!} accumulate on average 25\% of their mass after reddening; 
      at the same time, the top $\simeq$\,18\% of galaxies accumulate half or more of their $z=0$ stellar mass while on the red 
      sequence. The amount of $\Delta M_{\rm \star,red}$ is found to increase with increasing redshift zero $M_\star$. We 
      therefore advocate for an updated `mixed' scenario for the formation of the red population whereby galaxies become red 
      across a wide range of stellar mass, some fraction subsequently having $\Delta M_{\rm \star,red}=0$ while the remainder 
      grow by an amount $\Delta M_{\rm \star,red}>0$ which increases for more massive systems.
\end{itemize}

Together, the TNG100 and TNG300 simulations have a dynamic range of $10^7$ per spatial dimension. This enables us to connect 
the internal characteristics of galaxies -- notably, their colors -- to details of their host halos, as well as to the 
statistical properties of the global galaxy population within a full cosmological context. Our present investigation of the 
low redshift galaxy color distribution demonstrates the predictive and explanatory power of these simulations and their 
impressive utility as a theoretical tool for understanding the process of galaxy formation.


\section*{Acknowledgements}
DN would like to thank Charlie Conroy for development of the \textsc{FSPS} code and for helpful comments on the dust methods, 
Ben Johnson for development of the \textsc{prospector} code and assistance in its use, Vicente Rodriguez-Gomez for developing 
and allowing us to use the \textsc{SubLink} merger tree code, Arjen van der Wel for ideas on the spectral modeling, and 
Simon White for useful comments and suggestions.

VS, RW, and RP acknowledge support through the European Research Council under ERC-StG grant EXAGAL-308037 and would like to 
thank the Klaus Tschira Foundation. 
SG through the Flatiron Institute is supported by the Simons Foundation.
PT and SG are supported by NASA through Hubble Fellowship grants HST-HF-51384.001-A and HST-HF2-51341.001-A , respectively, awarded 
by the Space Telescope Science Institute, which is operated by the Association of Universities for Research in Astronomy, Inc., 
for NASA, under contract NAS5-26555.
MV acknowledges support from a MIT RSC award, the Alfred P. Sloan Foundation, and by NASA ATP grant NNX17AG29G. 
JN acknowledges support from NSF AARF award AST-1402480.

In the execution of the primary simulations presented herein, the authors gratefully acknowledge the Gauss Centre 
for Supercomputing (GCS) for providing computing time for the GCS Large-Scale Projects GCS-ILLU (2014) and GCS-DWAR (2016) on 
the GCS share of the supercomputer Hazel Hen at the High Performance Computing Center Stuttgart (HLRS). GCS is the alliance 
of the three national supercomputing centres HLRS (Universit{\"a}t Stuttgart), JSC (Forschungszentrum J{\"u}lich), and LRZ 
(Bayerische Akademie der Wissenschaften), funded by the German Federal Ministry of Education and Research (BMBF) 
and the German State Ministries for Research of Baden-W{\"u}rttemberg (MWK), Bayern (StMWFK) and Nordrhein-Westfalen (MIWF).
Additional simulations were carried out on the Hydra and Draco supercomputers at the Max Planck Computing and Data Facility
(MPCDF, formerly known as RZG), as well as on the Stampede supercomputer at the Texas Advanced Computing Center through 
the XSEDE project AST140063. Some additional computations in this paper were run on the Odyssey cluster supported by 
the FAS Division of Science, Research Computing Group at Harvard University.

\bibliographystyle{mnras}
\bibliography{refs}

\clearpage
\appendix

\section{The Dust Under the Rug} \label{sec_appendix}

In this appendix we show a few additional figures which explore the differences between the dust models and the 
resolution convergence of our results. 

\begin{figure}
\centerline{\includegraphics[angle=0,width=3.2in]{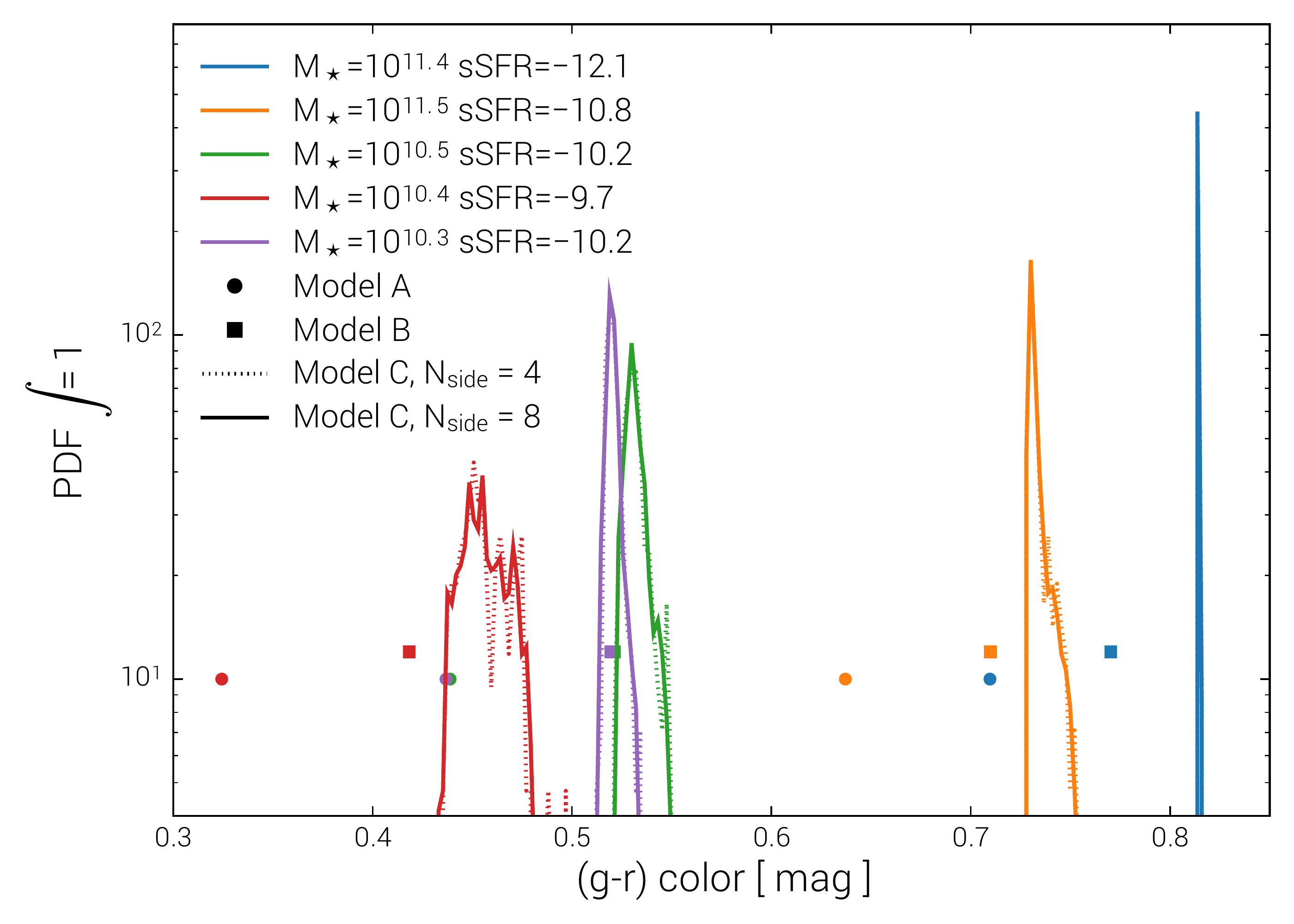}}
\caption{ Impact of viewing angle variation in the calculated (g-r) color values for five example systems spanning a range 
of stellar masses. For each galaxy, Model A and Model B results are shown by single circles or squares, respectively. The 
PDF of colors for Model C is then shown for each using high $N_{\rm side}$ parameters of 4 (dashed) or 8 (solid lines), 
corresponding to 192 and 768 projections per galaxy, respectively. The most massive galaxy has a predominantly spherical 
distribution of both stars and gas and thus exhibits little color variation due to projection-dependent dust effects. 
The less massive galaxies show asymmetric color distributions skewed to the red, although the total widths are 
typically less than 0.1 mag for all but the most dusty systems.
 \label{fig_viewing_angle_variation}} 
\end{figure}

First, Figure \ref{fig_viewing_angle_variation} shows the output of our color modeling on five individual galaxy examples 
spanning a range of stellar masses and star formation rates, as shown (colored lines). For each, the \mbox{(g-r)} colors 
derived with Model A and Model B are shown as single circles and squares, respectively. As these models have no viewing 
angle dependence, each galaxy has one unique color. The y-axis locations of these points are therefore arbitrary. PDFs of 
the Model C colors are also shown, for $N_{\rm side}=4$ (dotted; 192 projections per galaxy) and $N_{\rm side}=8$ (solid; 
768 projections per galaxy), the view-angle dependent color distributions converging with sufficient sampling of the sphere. 
Note that in the main text we have used $N_{\rm side}=1$ throughout, and taken either all 12 resultant colors per galaxy 
or a single random color of these twelve, depending on the specific application. 

\begin{figure}
\centerline{\includegraphics[angle=0,width=3.0in]{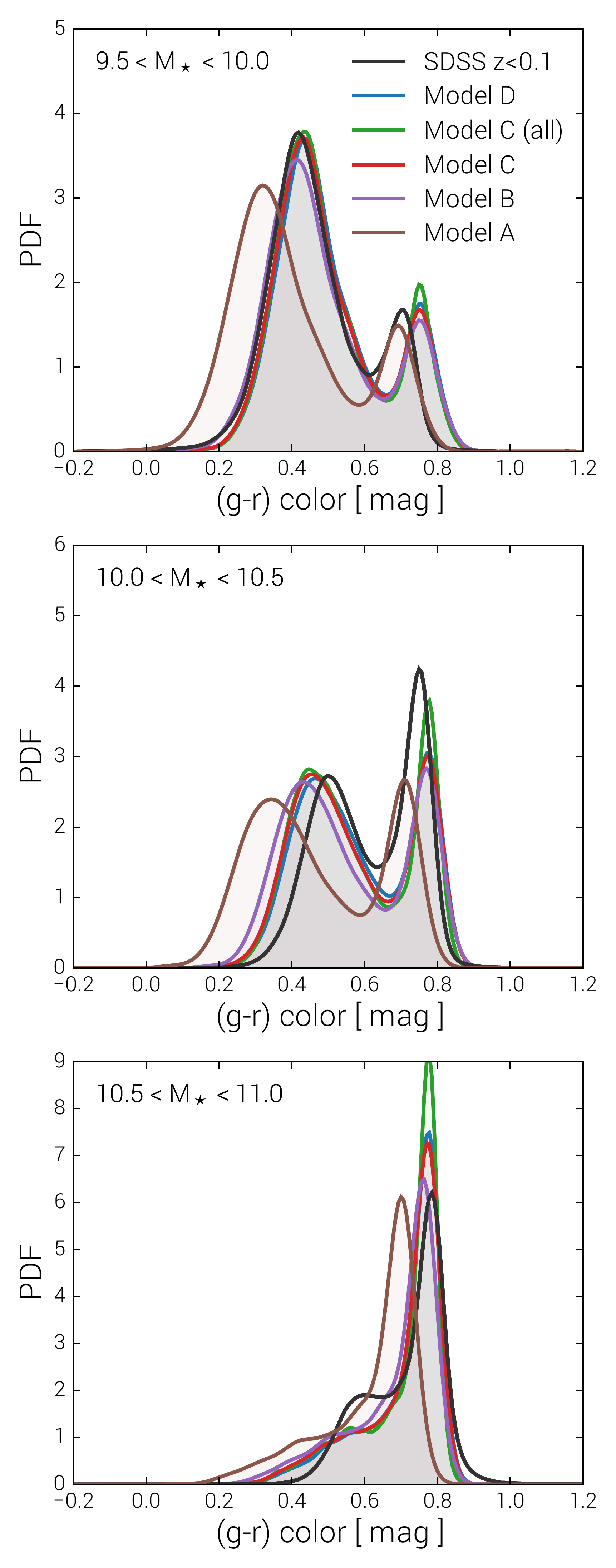}}
\caption{ Building up the dust model used in this paper, in the middle three 
stellar mass bins from $10^{9.5}$ $<$ $M_\star$/M$_{\rm sun}$ $<$ $10^{11.0}$. First we show Model A (brown lines), which includes 
the stellar population synthesis modeling with no dust treatment. Next, Model B (purple lines) includes only the unresolved 
presence of dense birth clouds. Model C  includes the resolved dust column calculations with one random line of sight per 
galaxy (red lines) or all twelve projections shown simultaneously (green lines). Model D uses a different dust geometry 
assumption (see text; blue lines).\label{fig_dust_model_comparison}} 
\end{figure}

Figure \ref{fig_dust_model_comparison} compares the different dust models on the full galaxy populations, in three stellar 
mass bins from TNG100-1. The unresolved dust correction of Model B is significant for low-mass (blue) galaxies, while the 
impact of Model C is relatively small in comparison. The modeled attenuation is also important in setting the precise location 
of the red peak for more massive galaxies, which is shifted blueward in the case of no dust. We note that Model C aims to 
account for attenuation by `resolved' gas structures in the simulation, and high density clumpy substructure in the ISM is 
far from resolved in the TNG simulations, implying that in higher resolution simulations we may be better able to model 
small-scale dust effects than in the current work. Including all twelve Model C colors for each galaxy or including only 
one at random has little impact on these global distributions, although this may not hold if we preferentially selected 
colors based on edge-on or face-on projections. 
We also include an as of yet unexplored Model D, a variation on Model C which considers an alternative 
geometrical assumption for the dust distribution along each line of sight. Specifically, it implements the `uniform scattering 
model' of \cite{calzetti94} instead of the internal dust model. In this case, the calculation of $\tau_\lambda$ is identical, 
and the attenuated luminosity is given by $L_{\rm obs}(\lambda) / L_{\rm i}(\lambda) = e^{-\tau_\lambda}$ instead of by 
Equation (\ref{eqn_model_C_lum}). The differences between Model C and Model D are small, and reflect only a minor change in 
the positive contribution of dust scattering to the emergent luminosity as a function of how the dust column is distributed 
along each line of sight through the gas distribution.

Finally, we comment on resolution convergence. As a reminder, we do not adjust in any way our model parameters as a function 
of resolution \citep[for details see][]{pillepich17a}. This is in contrast to the approach adopted in the EAGLE project where 
feedback mechanisms are specifically adjusted to counterbalance any (un)physical change with resolution and thereby to 
achieve a roughly, or at least more-so, constant result as a function of numerical resolution \citep{schaye15}. Different 
properties of our simulated galaxies have different convergence properties, the most fundamental of which are discussed in 
Appendix A of \cite{pillepich17a}. In Figure \ref{fig_stellar_ages_Z} we add to that discussion by showing the TNG100-1 and 
TNG300-1 predicted stellar ages and metallicities. In doing so, we are effectively looking at the resolution convergence of 
these quantities, since TNG300-1 is `one step lower' in resolution (a factor of two in spatial resolution, eight in mass) and 
therefore equivalent to the intermediate TNG100-2 run. We find that the stellar age and metallicity trends have somewhat 
different convergence behaviors. Given the low-statistics at the high-mass end for TNG100-1, the stellar ages 
appear to be well converged, better than most galaxy properties. Stellar metallicities increase, at fixed $M_\star$, 
with increasing resolution, by roughly $0.05 - 0.1$ dex between the TNG300-1 (intermediate) and TNG100-1 (high) 
resolution levels. We caution that, as in any case when showing $M_\star$ on the x-axis, any intrinsic shift in a specific 
galaxy property is convolved with the fact that TNG stellar masses are not perfectly converged, and so galaxies will also 
shift along the x-axis, typically towards higher $M_\star$ at fixed halo mass with increasing resolution.

\begin{figure}
\centerline{\includegraphics[angle=0,width=3.0in]{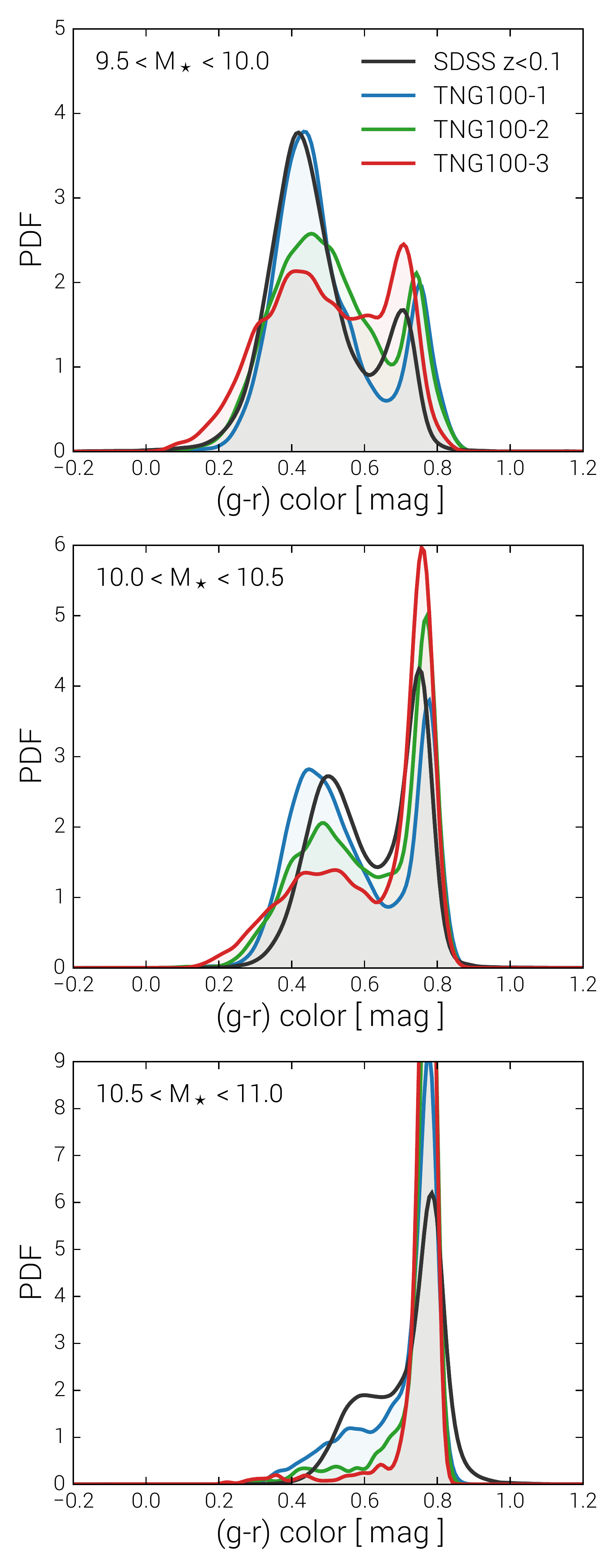}}
\caption{ The resolution convergence of the color distributions (Model C), shown in the central three stellar mass bins 
explored in this work, ranging from $10^{9.5}$ $<$ $M_\star$/M$_{\rm sun}$ $<$ $10^{11.0}$. Although the resolution 
convergence is fair, it is far from perfect. In general, with increasing mass and spatial resolution, galaxies are 
bluer, primarily as a result of larger stellar metallicities and larger stellar masses. Note however that galaxies also 
shift between bins due to changing stellar masses.\label{fig_res_convergence}} 
\end{figure}

Figure \ref{fig_res_convergence} presents the resolution convergence of the galaxy color PDFs directly, in the same three 
stellar mass bins as before. Keeping in mind that galaxies will shift between bins due to changing stellar masses, the 
general result is that with progressively lower resolution we lose low-mass blue galaxies. As a result, there are too many 
low mass red galaxies and the separation between the two populations becomes less well defined, and there is also an increase in 
the number of high-mass blue galaxies. 

{\renewcommand{\arraystretch}{1.2}
\begin{table*}
  \caption{Table of physical and numerical parameters for the three resolution levels of each of the two simulations 
  presented in this paper, TNG100 and TNG300. The physical parameters are: the box volume, the box side-length, the 
  initial number of gas cells, dark matter particles, and Monte Carlo tracer particles. The target baryon mass, 
  the dark matter particle mass, the $z$\,=\,0 Plummer equivalent gravitational softening of the collisionless 
  component, the same value in comoving units, the maximum softening applied to blackholes, and the minimum 
  comoving value of the adaptive gas gravitational softenings. Additional characterizations of the gas 
  resolution, measured at redshift zero: the minimum physical gas cell radius, the median gas cell radius, the 
  mean radius of star-forming gas cells, the mean hydrogen number density of star-forming gas cells, and the maximum 
  hydrogen gas density. The numerical parameters are: the number of high time frequency subbox volumes, the number of 
  snapshots in each subbox, the total number of timesteps to $z$\,=\,0, the total run time including substructure 
  identification in millions of CPU core hours, and the number of compute cores used.}
  \label{simTableBig}
  \begin{center}
    \begin{tabular}{lcllllll}
     \hline\hline
     
 Run Name & & TNG100-1 & TNG100-2 & TNG100-3 & TNG300-1 & TNG300-2 & TNG300-3 \\ \hline
 Volume & [\,cMpc$^3$\,] & $106.5^3$ & $106.5^3$ & $106.5^3$ & $302.6^3$ & $302.6^3$ & $302.6^3$ \\
 $L_{\rm box}$ & [\,cMpc/$h$\,] & 75 & 75 & 75 & 205 & 205 & 205 \\
 $N_{\rm GAS}$ & - & $1820^3$ & $910^3$ & $455^3$ & $2500^3$ & $1250^3$ & $625^3$ \\
 $N_{\rm DM}$ & - & $1820^3$ & $910^3$ & $455^3$ & $2500^3$ & $1250^3$ & $625^3$ \\
 $N_{\rm TRACER}$ & - & $2 \times 1820^3$ & $2 \times 910^3$ & $2 \times 455^3$ & $1 \times 2500^3$ & $1 \times 1250^3$ & $1 \times 625^3$ \\
 $m_{\rm baryon}$ & [\,M$_\odot$/$h$\,] & $9.4 \times 10^5$ & $7.6 \times 10^6$ & $6.0 \times 10^7$ & $7.6 \times 10^6$ & $5.9 \times 10^7$ & $4.8 \times 10^8$ \\
 $m_{\rm DM}$ & [\,M$_\odot / h$\,] & $5.1 \times 10^6$ & $4.0 \times 10^7$ & $3.2 \times 10^8$ & $4.0 \times 10^7$ & $3.2 \times 10^8$ & $2.5 \times 10^9$ \\
 $m_{\rm baryon}$ & [\,10$^6$\,M$_\odot$\,] & 1.4 & 11.2 & 89.2 & 11 & 88 & 703 \\
 $m_{\rm DM}$ & [\,10$^6$\,M$_\odot$\,] & 7.5 & 59.7 & 477.7 & 59 & 470 & 3764 \\
 $\epsilon_{\rm DM,stars}^{z=0}$ & [\,kpc\,] & 0.74 & 1.48 & 2.95 & 1.48 & 2.95 & 5.90 \\
 $\epsilon_{\rm DM,stars}$ & [\,ckpc/$h$\,] & 1.0 $\rightarrow$ 0.5 & 2.0 $\rightarrow$ 1.0 & 4.0 $\rightarrow$ 2.0 & 
                                              2.0 $\rightarrow$ 1.0 & 4.0 $\rightarrow$ 2.0 & 8.0 $\rightarrow$ 4.0 \\
 $\epsilon_{\rm BH,max}$ & [\,ckpc/$h$\,] & 5.0 & 5.84 & 6.84 & 5.84 & 6.84 & 13.7 \\
 $\epsilon_{\rm gas,min}$ & [\,ckpc/$h$\,] & 0.125 & 0.25 & 0.5 & 0.25 & 0.5 & 1.0 \\ \hline
 $r_{\rm cell,min}$ & [\,pc\,] & 14 & 74 & 260 & 47 & 120 & 519 \\
 $\bar{r}_{\rm cell}$ & [\,kpc\,] & 15.8 & 31.3 & 63.8 & 31.2 & 63.8 & 153.1 \\ 
 $\bar{r}_{\rm cell,SF}$ & [\,pc\,] & 355 & 720 & 1410 & 715 & 1420 & 3070 \\ 
 $\bar{n}_{\rm H,SF}$ & [\,cm$^{-3}$\,] & 1.0 & 0.6 & 0.5 & 0.6 & 0.5 & 0.4 \\ 
 $n_{\rm H,max}$ & [\,cm$^{-3}$\,] & 3040 & 185 & 30 & 490 & 235 & 30 \\ \hline
 $N_{\rm subboxes}$ & - & 2 & 2 & 2 & 3 & 3 & 3 \\
 $N_{\rm snaps,sub}$ & - & 7908 & 4380 & 2431 & 2449 & 3045 & 2050 \\
 $\Delta t$ & - & 11316834 & 3221234 & 500779 & 6203062 & 944143 & 209161 \\
 CPU Time & [\,Mh\,] & 18.0 & 0.6 & 0.02 & 34.9 & 1.3 & 0.05 \\
 $N_{\rm cores}$ & - & 10752 & 2688 & 336 & 24000 & 6000 & 768 \\
 \hline
 
    \end{tabular}
  \end{center}
\end{table*}}

{\renewcommand{\arraystretch}{1.2}
\begin{table*}
  \caption{Table of physical and numerical parameters as in Table \ref{simTableBig}, except for the six dark-matter only analog runs.}
  \label{simTableDM}
  \begin{center}
    \begin{tabular}{lcllllll}
     \hline\hline
     
  & & TNG100-1-DM & TNG100-2-DM & TNG100-3-DM & TNG300-1-DM & TNG300-2-DM & TNG300-3-DM \\ \hline
 Volume & [\,cMpc$^3$\,] & $106.5^3$ & $106.5^3$ & $106.5^3$ & $302.6^3$ & $302.6^3$ & $302.6^3$ \\
 $N_{\rm DM}$ & - & $1820^3$ & $910^3$ & $455^3$ & $2500^3$ & $1250^3$ & $625^3$ \\
 $m_{\rm DM}$ & [\,M$_\odot$\,] & $8.9 \times 10^6$ & $7.1 \times 10^7$ & $5.7 \times 10^8$ & $7.0 \times 10^7$ & $5.6 \times 10^8$ & $4.5 \times 10^9$ \\
 $\epsilon_{\rm DM}^{z=0}$ & [\,kpc\,] & 0.74 & 1.48 & 2.95 & 1.48 & 2.95 & 5.90 \\
 $\epsilon_{\rm DM}$ & [\,ckpc/$h$\,] & 1.0 $\rightarrow$ 0.5 & 2.0 $\rightarrow$ 1.0 & 4.0 $\rightarrow$ 2.0 & 
                                        2.0 $\rightarrow$ 1.0 & 4.0 $\rightarrow$ 2.0 & 8.0 $\rightarrow$ 4.0 \\
 \hline
    \end{tabular}
  \end{center}
\end{table*}}

Another culprit in the apparent lack of resolution convergence for the bluest systems arises directly from the finite mass 
resolution of the simulations and the stochastic sampling of the physical process of star formation. Namely, a low-mass 
galaxy on the star formation main sequence with a SFR of $\sim$\,1 \msun\!/yr will on average create only one 10$^6$\,\msun 
stellar particle every million years at TNG100-1 resolution, and only one 10$^7$\,\msun stellar particle every \textit{ten} 
million years at TNG100-2 resolution. This discreteness implies that at lower resolutions there may be no very young star 
particles present even though a galaxy is actively star forming. Motivated by this exact problem, \cite{trayford17} (using the 
EAGLE simulation) decided to implement a re-sampling technique in post-processing to effectively generate the expected populations 
of very young star particles. Indeed, even with perfect resolution convergence of the physical properties of a galaxy, its color 
would still become redder with worsening mass resolution due to such a stochastic sampling. This effect provides a caution for 
accurate colors in lower resolution simulations, and likely drives much of the resolution trend seen in this Figure.

Tables \ref{simTableBig} and \ref{simTableDM} give the most relevant physical and numerical parameters for the baryonic and 
dark-matter only runs, respectively. Each includes the highest resolution realization of the two boxes presented herein, 
TNG100-1 and TNG300-1, as well as the two lower resolution versions of each. Among other values we include the 
mean radius of star-forming gas cells $\bar{r}_{\rm gas,SF}$ in physical parsecs, and the mean number density of star-forming 
gas cells $\bar{n}_{\rm gas,SF}$ in particles per cm$^3$. These provide a rough idea of the spatial scales and gas densities 
at which the physical process of star formation takes place in our numerical simulations, namely, $\sim 350$\,pc and 
$\sim 1$\,cm$^{-3}$ in the case of TNG100-1.

\end{document}